\documentclass[prb,groupedaddress,amssymb,aps,showpacs,twocolumn]{revtex4}
\usepackage{graphicx,bm,amsmath}
\begin{document}

\title{Superconducting qubits coupled to nanoelectromechanical resonators: An architecture for solid-state quantum information processing}

\author{Michael R. Geller$^1$ and Andrew N. Cleland$^2$}

\affiliation{$^1$Department of Physics and Astronomy, University of Georgia,  Athens, Georgia 30602-2451 \\
$^2$Department of Physics and iQUEST, University of California, Santa Barbara, California 93106}

\date{September 25, 2004}

\begin{abstract}
We describe the design for a scalable, solid-state quantum-information-processing architecture based on the integration of GHz-frequency nanomechanical resonators with Josephson tunnel junctions, which has the potential for demonstrating a variety of single- and multi-qubit operations critical to quantum computation. The computational qubits are eigenstates of large-area, current-biased Josephson junctions, manipulated and measured using strobed external circuitry. Two or more of these {\it phase} qubits are capacitively coupled to a high-quality-factor piezoelectric nanoelectromechanical disk resonator, which forms the backbone of our architecture, and which enables coherent coupling of the qubits. The integrated system is analogous to one or more few-level atoms (the Josephson junction qubits) in an electromagnetic cavity (the nanomechanical resonator). However, unlike existing approaches using atoms in electromagnetic cavities, here we can individually tune the level spacing of the ``atoms'' and control their ``electromagnetic'' interaction strength. We show theoretically that quantum states prepared in a Josephson junction can be passed to the nanomechanical resonator and stored there, and then can be passed back to the original junction or transferred to another with high fidelity. The resonator can also be used to produce maximally entangled Bell states between a pair of Josephson junctions. Many such junction-resonator complexes can assembled in a hub-and-spoke layout, resulting in a large-scale quantum circuit. Our proposed architecture combines desirable features of both solid-state and cavity quantum electrodynamics approaches, and could make quantum information processing possible in a scalable, solid-state environment.
\end{abstract}

\pacs{03.67.Lx, 85.25.Cp, 85.85.+j}

\maketitle
\clearpage

\section{INTRODUCTION\label{introduction section}}

The lack of a large collection of physical qubit elements, having both sufficiently long quantum-coherence lifetimes and the means for producing and controlling their entanglement, remains the principal roadblock to building a large-scale quantum computer. Superconducting devices have been understood for several years to be natural candidates for quantum computation, given that they exhibit robust macroscopic quantum behavior.\cite{MakhlinRMP01} Demonstrations of long-lived Rabi oscillations in current-biased Josephson tunnel junctions,\cite{YuSci02,MartinisPRL02} and of both Rabi oscillations and Ramsey fringes in a Cooper-pair box,\cite{NakamuraNat99,NakamuraPRL02,VionSci02} have generated significant new interest in the potential for
superconductor--based quantum computation.\cite{LeggettSci02,Search&DiscoveryJune02} Several additional experimental accomplishments have followed,\cite{PashkinNat03,ChiorescuSci03,BerkleySci03,YamamotoNat03,CrankshawPRB04,SimmondsPRL04,WallraffNat04,CooperPre04} including the impressive demonstration of controlled-NOT logic with charge qubits,\cite{YamamotoNat03} and a large body of theoretical work is beginning to address these and related systems.\cite{ShnirmanPRL97,MakhlinNat99,MooijSci99,MakhlinJLTP00,YouPRL02,Yukon02,SmirnovPre02,BlaisPRL03,PlastinaPRB03,ZhouPRA04,Buisson01,BlaisPRA04,WallraffNat04,GirvinPre03,MarquardtPRB01,HekkingPre02,ZhuPRA03,ArmourPRL02,Armour02,IrishPRB03,IoffeNat99,BlaisPRA00,PlastinaPRB01,SiewertPRL01,IoffeNat02,TianPRB02,PaladinoPRL02,IoffePRB02,FaoroPRL03,JohnsonPRB03,ZazunovPRL03,MartinisPRB03,KimPRB03,StrauchPRL03,SteffenPRB03,Cleland&GellerPRL04} Coherence times $\tau_{\rm \varphi}$ up to 5$\, \mu$s have been reported in the current-biased devices,\cite{YuSci02} with corresponding quantum-coherent quality factors $Q_{\rm \varphi} \equiv \tau_{\rm \varphi} \, \Delta E / \hbar$ of the order of $10^5 \! ,$ indicating that these systems should be able to perform many logical operations during the available coherence lifetime.\cite{operationsnote} Here $\Delta E$ is the qubit energy-level separation, which was $68 \, \mu {\rm eV}$ in the experiment of Ref.~[\onlinecite{YuSci02}].

In this paper, we expand on our earlier proposal suggesting that GHz-frequency nanoelectromechanical resonators can be used to coherently couple two or more  current-biased Josephson junction (JJ) devices together to make a flexible and scalable solid-state quantum-information-processing architecture.\cite{Cleland&GellerPRL04} The computational qubits are taken to be the energy eigenstates of the JJs, which are to be individually prepared, controlled, and measured using the external circuitry developed by Martinis {\it et al.}\cite{MartinisPRL02} These superconducting {\it phase} qubits are capacitively coupled to a high-quality-factor piezoelectric dilatational disk resonator, cooled on a dilution refrigerator to the quantum limit, which forms the backbone of our architecture. We shall show that the integrated system is analogous to one or more few-level atoms (the JJs) in an electromagnetic cavity (the resonator). However, here we can individually tune the energy level spacing of each ``atom", and control the ``electromagnetic" interaction strength. This analogy makes it clear that our design is sufficiently flexible to be able to carry out essentially any operation that can be done using other architectures, provided that there is enough coherence. Many of our results will apply to other architectures that are similar to atoms in a cavity.

Several investigators have proposed the use of $LC$ resonators,\cite{ShnirmanPRL97,MakhlinNat99,MooijSci99,MakhlinJLTP00,YouPRL02,Yukon02,SmirnovPre02,BlaisPRL03,PlastinaPRB03,ZhouPRA04}
superconducting cavities,\cite{Buisson01,BlaisPRA04,WallraffNat04,GirvinPre03} or other types of oscillators,\cite{MarquardtPRB01,HekkingPre02,ZhuPRA03} to couple JJs together. We note that although harmonic oscillators are ineffective as computational qubits, because the lowest pair of levels cannot be frequency selected by an external driving force, they are quite desirable as bus qubits or coupling elements. Early on, Shnirman {\it et al.} \cite{ShnirmanPRL97} suggested an architecture consisting of several superconducting charge qubits in parallel with an inductor. The JJs are themselves out of resonance with each other and hence weakly coupled, and the resulting $LC$ resonator (the capacitance coming from the junction geometry) is also used well below its resonant frequency. An interesting modification of this design couples the small island to the external circuit through a pair of parallel JJs, which allows the Josephson coupling energy to be varied, using an external magnetic field.\cite{MakhlinNat99} To date, however, the only coupled superconducting qubits demonstrated experimentally have been the capacitively coupled charge qubits of Peshkin {\it et al.}\cite{PashkinNat03} and Yamamoto {\it et al.},\cite{YamamotoNat03} and the capacitively coupled phase qubits of Berkley {\it et al.}\cite{BerkleySci03} 

Resonator-based coupling schemes, such as the one proposed here, have the advantage of additional functionality resulting from the ability to tune the qubits relative to the resonator frequency, as well as to each other. We shall show that by tuning the JJs in and out of resonance with the nanomechanical  resonator, qubit states prepared in a junction can be passed to the resonator and stored there, and can later be passed back to the original junction or transferred to another JJ with high fidelity. The resonator can also be used to produce controlled entangled states between a pair of JJs. Alternatively, when both qubits are detuned from the resonator, the resonator produces a weak (higher order) ``dispersive" qubit coupling similar to that of a capacitor. The use of mechanical resonators to mediate multi-qubit operations in JJ--based quantum information processors has not (to the best of our knowledge) been considered previously, but our proposal builds on the interesting recent theoretical work by Armour {\it et al.}\cite{ArmourPRL02,Armour02} and Irish {\it et al.}\cite{IrishPRB03} on the entanglement of a nanoelectromechanical resonator with a single Cooper-pair box. In fact, there is currently a big effort to push a variety of nanomechanical systems to the quantum limit.\cite{KnobelNat03,LaHayeSci04,Blencowe04}

In the next section we recall the basic properties of large-area, current-biased JJs. In Sec.~\ref{model section} we discuss our proposed architecture, and construct a simple model Hamiltonian to describe it. State-preparation and readout have been described elsewhere and will only be discussed briefly. The properties of the nanomechanical resonator are also described here in detail. In the remainder of the paper we discuss a variety of elementary single- and multi-qubit operations central to quantum computation: In Sec.~\ref{qubit storage and transfer section} we show how a qubit state prepared in a JJ can be passed to the nanomechanical resonator, stored there coherently, and later passed back to the original junction or transferred to another JJ. Two-junction entanglement, mediated by the resonator, is studied in Sec.~\ref{entanglement section}. In
Sec.~\ref{network section} we show how our architecture can be extended to make a large-scale quantum circuit. Our conclusions are given in Sec.~\ref{discussion section}. Several immediate extensions of the present work, including the development of protocols for universal two-qubit quantum logic, are currently in progress and will be discussed in future publications.

\section{THE CURRENT-BIASED JOSEPHSON JUNCTION\label{junction section}}

%\begin{figure}
%\includegraphics[width=.45\textwidth]{fig1.jpg}
%\caption[]{\label{fig.washboard}(main panel) Effective potential $U(\delta)$ for dimensionless bias current $s \equiv I_b/I_0$ equal to $0.1,$ plotted in units of $E_{\rm J}$. (inset) Equivalent-circuit model for a current-biased Josephson junction. A capacitance $C$ and resistance $R$ are in parallel with an ``ideal" Josephson element, represented by a cross and having critical current $I_0$. A bias current $I_b$ is driven through the circuit. }
%\end{figure}

Our architecture relies on the use of large-area JJs, biased with a current $I_{\rm b}$, which can be quasi-static or have oscillatory components. The junctions have a large capacitance $C$ (typically 1 to $50 \, {\rm pF}$) and critical current $I_0$ (in the 10 to $150 \, {\mu \rm A}$ range) so that the largest relevant energy scale in the system is the Josephson coupling energy
\begin{equation}
E_{\rm J} \equiv {\hbar I_0 \over 2 e},
\end{equation}
where $e$ is the magnitude of the electron charge. In contrast, the Cooper-pair charging energy
\begin{equation}
E_{\rm c} \equiv {(2e)^2 \over 2C}
\end{equation}
is small compared with $E_{\rm J}$, and is also usually smaller than the thermal energy $k_{\rm B} T$. For example,
\begin{equation}
E_{\rm J} =  2.05 \, {\rm meV} \! \times \!  I_0[{\rm \mu A}]  
\ \ \ \ {\rm and} \ \ \ \ 
E_{\rm c} = {320 \, {\rm neV}  \over C[{\rm pF}]},
\end{equation}
where $I_0[{\rm \mu A}]$ and $C[{\rm pF}]$ are the critical current and junction capacitance in microamperes and picofarads, respectively.

\subsection{Semiclassical junction dynamics\label{semiclassical section}}

The dynamics of a real JJ can be understood as following from the equivalent circuit model shown in the inset to Fig.~\ref{fig.washboard}, known as the resistively and capacitively shunted junction model,\cite{StewartAPL68,McCumberJAP68} where the ``ideal'' Josephson element controls the superconducting component $I_{\rm s}$ of the total electrical current $ I_{\rm s} + I_{\rm n}$ in accordance with the well-known Josephson equations
\begin{equation} 
I_{\rm s} =I_0 \sin \delta \ \ \ \ {\rm and} \ \ \ \  {d \delta \over dt} = {2 e V \over  \hbar}.
\label{josephson equations}
\end{equation}
Here $I_{\rm s}$ is the supercurrent flowing through the ideal Josephson junction element, $\delta$ is the difference between the phases of the (spatially uniform) superconducting order parameters on each side of the junction, and $V$ is the voltage across the junction.  $I_{\rm n}$ is the normal component of the current. The resistance $R$ accounts for finite-temperature quasiparticle tunneling as well as electron tunneling in the finite-voltage state. Equating  the sum of the currents flowing through the capacitor, ideal junction, and resistor, to $I_{\rm b}$, leads to 
\begin{equation}
{\hbar^2 \over 2 E_{\rm c}} \, {d^2 \delta \over dt^2} + {\hbar^2 \over 4 e^2 R} \, {d \delta \over dt} 
+ E_{\rm J} \big(\sin \delta - s \big) = 0,
\label{primitive equation of motion}
\end{equation}
where $s \equiv I_{\rm b}/I_0$ is the dimensionless bias current. By rewriting Eq.~(\ref{primitive equation of motion}) in the equivalent form
\begin{equation}
M {d^2 \delta \over dt^2} = - {dU \over d \delta} - \eta {d \delta \over dt},
\label{equation of motion}
\end{equation}
it can be interpreted as the equation of motion for a particle of ``mass'' 
\begin{equation}
M \equiv {\hbar^2 \over 2 E_{\rm c}}
\label{effective mass}
\end{equation}
moving in an effective potential
\begin{equation}
U(\delta) \equiv - E_{\rm J} \big(\cos \delta + s \, \delta \big),
\label{effective potential}
\end{equation}
and in the presence of velocity-dependent dissipation characterized by a friction coefficient $\eta \equiv \hbar^2 /4 e^2 R$. Note that $M$ actually has dimensions of mass $\times$ length$^2 \! .$

The potential $U(\delta)$, which resembles a tilted washboard, is shown in the main panel of Fig.~\ref{fig.washboard} for a dimensionless bias current of $s = 0.1$. The zero-voltage state of the junction corresponds to the particle or phase variable being trapped in one of the metastable minima present when $s < 1$, and the finite-voltage state corresponds to the phase variable running down the washboard potential. In what follows we will assume $0 \le s < 1,$ and without loss of generality we can also assume that $0 \le \delta < 2 \pi$.

The potential $U(\delta)$ reaches its minimum and maximum values in the domain $0 \le \delta < 2 \pi$ at $\delta_{\rm min} = \arcsin s$ and  $\delta_{\rm max} = \pi - \arcsin s$. The depth $\Delta U \equiv U(\delta_{\rm max}) - U(\delta_{\rm min})$ of the potential well is
\begin{equation}
\Delta U = 2 E_{\rm J} \left [ \sqrt{1-s^2} - s \arccos s \right ],
\label{well depth}
\end{equation}
which vanishes as
\begin{equation}
\Delta U \rightarrow \frac{4\sqrt{2}}{3} E_{\rm J} (1-s)^{3/2}
\end{equation}
in the $s \rightarrow 1^-$ limit.

The curvature $U''(\delta)$ at the minimum of the potential is used to define the junction's plasma frequency, 
\begin{equation}
\omega_{\rm p} \equiv \sqrt{U''(\delta_{\rm min}) \over M} = \omega_{{\rm p}0} \big(1-s^2\big)^{1/4} \! ,
\label{eq.plasmafreq}
\end{equation}
which is the frequency of small oscillations of $\delta$ about $\delta_{\rm min}$. Here $M$ is the effective mass defined in Eq.~(\ref{effective mass}), and
\begin{equation}
\omega_{{\rm p}0} = \sqrt{2 e I_0 \over \hbar C} = {\sqrt{2 E_{\rm c} E_{\rm J}} \over \hbar}
\label{zero-bias frequency}
\end{equation}
is the plasma frequency at zero bias.

The dependence of the barrier height and plasma frequency on bias current are plotted in Fig.~\ref{fig.plasmabarrier}. For junctions appropriate for quantum computation, $\omega_{{\rm p}0} /2 \pi$ is typically in the range of 5 to $50 \, {\rm GHz}$. The barrier height during state preparation and readout is usually adjusted so that $\Delta U / \hbar  \omega_{\rm p}$ is between 3 and 5, but, as we shall discuss below, is it advantageous to keep $s$ smaller during actual quantum computation.

%\begin{figure}
%\includegraphics[width=.4\textwidth]{fig2.jpg}
%\caption[]{\label{fig.plasmabarrier}Barrier height and plasma frequency as a function of the dimensionless bias current $s$. Here $\Delta U_0 \equiv 2 E_{\rm J}$ is the barrier height at zero bias, and $\omega_{{\rm p}0}$ is the zero-bias plasma frequency defined in Eq.~(\ref{zero-bias frequency}).}
%\end{figure}

The effect of dissipation, caused in the resistively and capacitively shunted junction model by the resistance $R$, can be characterized by the number of oscillations at the plasma frequency during the relaxation time $RC$, or $\omega_{\rm p} RC$. In what follows we will assume that dynamics is highly underdamped, with $\omega_{\rm p} RC \gg 1$.

\subsection{Quantizing the low-energy junction dynamics: The phase qubit\label{quantization section}}

When the thermal energy $k_{\rm B} T$ and energy decay width $\hbar / RC$ are both smaller than $\hbar \omega_{\rm p}$, quantum fluctuations of $\delta$ become important, and the JJ has to be treated quantum mechanically. This limit was studied in the 1980's as an example of a single macroscopic degree of freedom---the difference between phases of order parameters---that nonetheless behaves quantum mechanically.\cite{Caldeira83,MartinisPRL85,MartinisPRB87,ClarkeSci88} This is also the regime of current interest for applications to quantum computing.

When dissipation is absent, the low-energy dynamics can be quantized by introducing a Lagrangian $L_{\rm J} = {1 \over 2}M  {\dot \delta}^2 - U$ and canonical momentum $P \equiv \partial L / \partial {\dot \delta} = M {\dot \delta}$ associated with the $\eta = 0$ limit of Eq.~(\ref{equation of motion}). According to the Josephson equations, $P$ is proportional to the charge ${\sf Q}$ or to the number of Cooper pairs ${\sf Q}/2e$ on the capacitor according to $P = \hbar {\sf Q}/2e$. The classical Hamiltonian is $P^2/2M + U$. To quantize the system, we let $P = -i \hbar {d \over d \delta}$, so that $[\delta , P ] = i \hbar.$ Then the quantized Hamiltonian is
\begin{equation}
H_{\rm J} = - E_{\rm c} {d^2 \over d \delta^2} + U(\delta),
\label{junction hamiltonian}
\end{equation}
and the dynamics is governed by the Schr\"odinger equation $ i \hbar \partial_t \psi  =  H_{\rm J} \psi$. Because $U$ depends on $s$, which itself depends on $t$, $H_{\rm J}$ is generally time-dependent.

Naively, the stationary states and energies of the JJ with fixed $s$ follow straightforwardly from the  one-dimensional eigenvalue problem
\begin{equation}
H_{\rm J}  \, \psi_m(\delta) = \epsilon_m \, \psi_m(\delta), \ \ \ \ \ m = 0, 1, 2, \dots
\label{stationary states}
\end{equation}
However, a careful analysis\cite{ZwergerPRB86} shows that in the presence of any finite ohmic dissipation (finite $\eta$), quantum coherence between the different wells in $U(\delta)$ is destroyed. This, in fact, justifies the use of the washboard potential in the first place: Strictly speaking, $\delta$ is a periodic variable,
with $\delta$ physically equivalent to $\delta + 2 \pi$. In what follows, we will work with stationary states associated with a single potential minimum (in the domain $0 \le \delta < 2 \pi$). It is these stationary states that are of interest to quantum computation.

When $s=0$, the junction contains many (of order $\sqrt{E_{\rm J}/E_{\rm c}}$) bound states, the lowest of which are like that of a harmonic oscillator with level spacing $\hbar \omega_{{\rm p}0}$. The uniform spacing of the low-lying levels makes them difficult to address individually with a classical external driving force. Therefore, state preparation is carried out with $s$ just below unity, in which case there are only a few quasibound states $|0\rangle, \, |1\rangle, \, |2\rangle, \dots$ present, and the effective potential $U(\delta)$ becomes anharmonic and approximately cubic, as illustrated in Fig.~\ref{fig.currentbiasjj}. The remarkable 1985 spectroscopic observation\cite{MartinisPRL85} of these quantized states provided the first clear evidence for the quantum behavior of the macroscopic phase-difference variable $\delta$.

%\begin{figure}
%\includegraphics[width=.35\textwidth]{fig3.jpg}
%\caption[]{\label{fig.currentbiasjj}Metastable potential well in the cubic limit, showing the barrier of height $\Delta U$ that separates the metastable states $| 0 \rangle$, $| 1 \rangle$, and $| 2 \rangle$, from the continuum. This figure applies to the case of bias currents $s$ just below 1. The lowest two states are separated in energy by $\Delta E$.}
%\end{figure}

The lowest two eigenstates, $|0\rangle$ and $|1\rangle$, define a {\it phase} qubit. As stated, in the $s \lesssim1$ limit the potential is anharmonic, and the qubit level spacing
\begin{equation}
\Delta E \equiv \epsilon_1 - \epsilon_0
\label{level spacing}
\end{equation}
is somewhat smaller than $\hbar \omega_{\rm p}$, where $\omega_{\rm p}$ is the $s$-dependent plasma frequency. 

The qubit state is also usually measured with $s$ just below unity: In the absence of thermal or quantum fluctuations, switching to the finite-voltage state occurs when the bias current exceeds $I_0$. However, in a real junction, the finite-voltage state will occur before $I_{\rm b}$ reaches $I_0$, either because of thermal activation over the barrier or by quantum tunneling through it. Once the phase variable escapes into the continuum, it runs down the corrugated potential, and a voltage $V$ of approximately $2 \Delta_{\rm sc}/e$ develops across the junction, where $\Delta_{\rm sc}$ is the superconducting energy gap ($\Delta_{\rm sc} \! \approx 180 \, \mu {\rm eV}$ for Al junctions). The supercurrent component then oscillates with angular frequency $2e V/\hbar$---the AC Josephson effect. The thermal activation regime has been explored in detail, for various limits of dissipation.\cite{FultonPRB74,DevoretPRL84,DevoretPRB87} For fixed current bias, the thermal activation rate falls exponentially with inverse temperature, until the dominant escape mechanism becomes quantum tunneling.\cite{DevoretPRL85,Caldeira83,denBoer80,JackelPRL81,MartinisPRB87}  At temperatures low enough so that quantum tunneling dominates thermal activation, the qubit state can be observed by measuring the tunneling rate, which is strongly state-dependent. State preparation and readout are discussed further in Sec.~\ref{preparation and readout section}.

The barrier height $\Delta U$ and the energy splitting $\Delta E$ (through its dependence on $\omega_{\rm p}$) are both strong functions of the bias current $s$. The ability to tune the plasma frequency is one of the current-biased Josephson junction's great strengths and weaknesses. It enables the qubit level-spacing  $\Delta E$ to be tuned adiabatically into resonance with another qubit or, as in our approach, with a resonator, but it also makes the circuit sensitive to bias-current noise, as characterized by the non-zero derivative $d\omega_{\rm p} /d s$. Fluctuations in $s$ will generate noise and hence decoherence in the JJ. Although current methods of state preparation and measurement require $s$ very close to unity (typically near 0.99), where $d\omega_{\rm p} /d s$ is unfortunately large, the information-processing operations we describe below do {\it not}. In our simulations, we find it convenient to work with $s$ below $0.90$. 

\begin{table}
\caption{\label{energy table}
Energies $\epsilon_m$ of low-lying eigenstates as a function of dimensionless bias current $s$, for the JJ investigated in Ref.~[\onlinecite{MartinisPRL02}], with parameters $I_0 = 21 \, \mu {\rm A} \ (E_{\rm J} = 43.05 \, {\rm meV})$ and $C = 6 \, {\rm pF} \ (E_{\rm c} = 53.33 \, {\rm neV})$. Energies below are given in units of $\hbar \omega_{\rm p}$ and are measured relative to $U(\delta_{\rm min})$.  All dissipation and decoherence effects are neglected. The first column, labeled by $m+{1\over 2},$ gives the energies of the corresponding harmonic oscillator eigenfunctions, which are found to be extremely accurate for small $s$.}
\begin{ruledtabular}
\begin{tabular}{|c|cccc|}
junction state $|m\rangle$ & $m+{1\over2}$ & $s=0.50$ & $s=0.70$ & $s=0.90$ \\ \hline
$m=0$   &  0.500 & 0.500  & 0.500 & 0.500 \\
$m=1$   & 1.500 &  1.500 & 1.499 & 1.497 \\
$m=2$   & 2.500 &  2.499 & 2.498 & 2.492 \\
$m=3$   & 3.500 &  3.498 & 3.496 &  3.485
\end{tabular}
\end{ruledtabular}
\end{table}

The energies $\epsilon_m $ of the lowest four JJ states of the device used in Ref.~[\onlinecite{MartinisPRL02}], for a range of bias currents, are given in Table \ref{energy table} in units of $\hbar \omega_{\rm p}.$ We calculate these energies numerically by diagonalizing the Hamiltonian $H_{\rm J}$ of Eq.~(\ref{junction hamiltonian}) in a basis of harmonic oscillator eigenfunctions
\begin{equation}
\phi_m \equiv (2^m m! \sqrt{\pi} \, \ell_s)^{-1/2} \, e^{-\xi^2/2} \, H_m(\xi)
\label{oscillator basis functions}
\end{equation}
that are constructed by making a quadratic approximation
\begin{equation}
U(\delta) \approx U(\delta_{\rm min}) + {1 \over 2} \, U''(\delta_{\rm min}) \big(\delta - \delta_{\rm min}\big)^2
\end{equation} 
to $U(\delta)$ about its minimum at $\delta_{\rm min} = \arcsin s$. The $H_m, \ m \! = \! 0,1,2,\dots,$ are Hermite polynomials, and  $\xi \equiv (\delta - \delta_{\rm min}) / \ell_s$ is a centered and scaled phase variable, with
\begin{equation}
\ell_s \equiv \sqrt{ \hbar \over M \omega_{\rm p}} = \bigg({2 E_{\rm c} \over E_{\rm J}}\bigg)^{\! {1\over4}}
 \! \big(1 - s^2 \big)^{-{1\over8}}
\label{length scale}
\end{equation}
giving the characteristic width in $\delta$ of these eigenfunctions. We find rapid convergence to the values reported in Table \ref{energy table} as the number of harmonic oscillator basis states is increased to include all basis states with energies less than $U(\delta_{\rm max})$.

\begin{table}
\caption{\label{dipole table}
Dipole moments $x_{m m'}$ between pairs of low-lying JJ eigenstates states for bias $s = 0.90$. The entries with dots follow from symmetry. Junction parameters are the same as in Table \ref{energy table}.}
\begin{ruledtabular}
\begin{tabular}{|c|cccc|}
$\langle m |\delta | m' \rangle$ & $m'=0$ & $m'=1$ & $m'=2$ & $m'=3$ \\ \hline
$m=0$    &  1.12 & $3.46 \! \times \! 10^{-2}$  & $-5.86 \! \times \! 10^{-4}$ & $7.09 \! \times \! 10^{-6}$ \\
$m=1$    & $\cdot$ &  1.12 & $4.89 \! \times \! 10^{-2}$ & $-1.02 \! \times \! 10^{-3}$ \\
$m=2$    & $\cdot$ & $\cdot$  & 1.13 &  $6.00 \! \times \! 10^{-2}$ \\
$m=3$   & $\cdot$ & $\cdot$ & $\cdot$ &  1.13
\end{tabular}
\end{ruledtabular}
\end{table}

Dipole-moment matrix elements
\begin{equation}
x_{m m'} \equiv \langle m |\delta | m' \rangle,
\label{dipole moment definition}
\end{equation} 
which will also be used below, are calculated at bias $s = 0.90$ for the junction used in Ref.~[\onlinecite{MartinisPRL02}], using this same method. The results are given in Table \ref{dipole table}. All basis functions with energies less than $U(\delta_{\rm max})$ are included, and the oscillator-strength sum rules (adapted for this Hamiltionian) are satisfied to better than 99.999\%. Because the eigenfunctions are real, the matrix $x_{mm'}$ is symmetric, and with an appropriate choice of overall signs of the eigenfunctions, the first band of off-diagonal matrix elements can be made positive. The diagonal elements are also positive here, a consequence of our restriction to the domain $0 \le \delta < 2 \pi$.

The diagonal elements $x_{mm}$ are very close to $\delta_{\rm min},$ regardless of $m.$ In the $s \! = \! 0.90$ case considered in Table \ref{dipole table}, $\delta_{\rm min}$ is about 1.120. The values of off-diagonal elements of the form $x_{m,m \pm 1}$ can be understood by noting that for harmonic oscillator states, which in this case are close to the exact eigenfunctions,
\begin{equation}
\int \! d\delta \ \phi_m(\delta) \, \delta \, \phi_{m+1}(\delta) = \sqrt{m+1 \over 2} \ \ell_s
\label{dipole estimate}
\end{equation}
with $\ell_s = 4.883 \! \times \! 10^{-2}$. The remaining off-diagonal elements, which result from the small mixing of the harmonic oscillator states, are smaller than these by at least an order of magnitude.

\section{ARCHITECTURE AND MODEL HAMILTONIAN\label{model section}}

We turn now to the main focus of our paper, the description of a solid-state quantum-information-processing architecture consisting of a network of current-biased Josephson junctions coupled to nanoelectromechanical resonators. We will first consider a single nanomechanical resonator coupled to one or two JJ qubits; the extension to larger systems will be considered below in Sec.~\ref{network section}, as well as in future work. 

\begin{widetext}

%\begin{figure}
%\includegraphics[width=0.95\textwidth]{fig4.jpg}
%\caption[]{\label{fig.twoqubit}  Two-qubit circuit diagram. The computational qubits are the two JJs in the center, shown as crossed boxes, each coupled to one side of the piezoelectric disk resonator. Each crossed box represents a real JJ, modeled by an ideal Josephson element in parallel with a resistor and capacitor. The current bias and readout circuits for each qubit circuit are shown on the left and right sides of the figure. Note that there is no direct electrical connection between the two qubits.}
%\end{figure}

\end{widetext}

The complete circuit diagram for the two-JJ circuit is shown in Fig.~\ref{fig.twoqubit}. The two central crossed boxes  are the JJs to be used as phase qubits, and they include the parallel capacitance and resistance shown in the inset to Fig.~\ref{fig.washboard}. The disk-shaped element in the center of the figure is the nanomechanical resonator, consisting of a single-crystal piezoelectric disk sandwiched between two metal electrodes. Applying a voltage across this element produces an electric field between the plates, and through the piezoelectric response, a strain in the crystal. Conversely, strain in the resonator produces a charge on the electrodes, whose rate of change contributes to the current flowing through a JJ. 

\subsection{Single-qubit state preparation, manipulation, and readout\label{preparation and readout section}}

Two of the most critical factors in the design of a successful JJ-based quantum information processor are high-impedance bias and high-fidelity readout circuits that do not disturb the qubit during computation. This is currently a subject of active experimental investigation, and for concreteness we will assume the bias circuit design developed recently by Martinis {\it et al.},\cite{MartinisPRL02} but we will leave the the readout circuitry unspecified. Our architecture can be adapted to improved readout schemes as they become available.

State preparation and readout are performed with $I_{\rm b}$ just below $I_0$, where $U(\delta)$ is anharmonic and shallow. The anharmonicity allows preparation from a harmonically varying bias current, which is tuned to couple to only the lowest two states. The $|0\rangle$ state is prepared by waiting for any excited component to decay. The state $|1\rangle$, or a superposition $\alpha |0\rangle + \beta |1\rangle$, is prepared by adding radiofrequency (RF) components of magnitudes $I_{\rm rf}^{\rm c}$ and $I_{\rm rf}^{\rm s}$ to the DC bias current, in the form\cite{MartinisPRB03}
\begin{equation}
I_{\rm b}(t) = I_{\rm dc} + I_{\rm rf}^{\rm c} \cos(\omega_{\rm rf}t) + I_{\rm rf}^{\rm s} \sin(\omega_{\rm rf} t),
\end{equation}
with $I_{\rm dc}$ and $I_{\rm rf}^{\rm s,c}$ all varying adiabatically (slow compared with the frequency $\Delta E/\hbar$). When $\omega_{\rm rf}$ is nearly resonant with $
\Delta E / \hbar$, the qubit will undergo Rabi oscillations, allowing the preparation of arbitrary linear combinations of $|0\rangle$ and $|1\rangle$. The associated Rabi frequency
\begin{equation}
\Omega_{\rm rf} \equiv \sqrt{ \bigg( {s_{\rm rf} x_{01} E_{\rm J} \over \hbar}\bigg)^2 
+ \bigg( \omega_{\rm rf}  - {\Delta E \over \hbar} \bigg)^2 } 
\label{RF Rabi frequency}
\end{equation}
depends on both $s_{\rm rf} \equiv I_{\rm rf} / I_0$ and the detuning. All states on the Bloch sphere may be prepared in this manner.\cite{MartinisPRB03}

Readout of a JJ state $\alpha |0\rangle + \beta |1\rangle$ is performed by then tuning $\omega_{\rm rf}$ into resonance with $(\epsilon_2 - \epsilon_1)/\hbar$, thereby exciting the qubit component in the $|1\rangle$ state up to $|2\rangle$, out of which it quickly tunnels, thereby resulting in a measurement of $|\beta|$. Martinis {\it et al.}\cite{MartinisPRL02} have established that a single-shot readout of the JJ states $|0\rangle$ and $|1\rangle$ can be performed with 99\%  and 85\% accuracy, respectively.

\subsection{Nanomechanical resonator\label{resonator section}}

The second important element in our design is the use of piezoelectric nanoelectromechanical disk resonators, with dilatational-mode frequencies $\omega_0/2 \pi$ in the $1 \ {\rm to} \  50 \, {\rm GHz}$ range. Piezoelectric dilatational resonators with frequencies in this range, and quality factors $Q \equiv \omega_0 \tau$ of the order of $10^3$ at room temperature, have been fabricated from sputtered AlN.\cite{Ruby94,Ruby01} Here $\tau$ is the energy damping time. The radius of the disk is denoted by $R$, and $b$ is its thickness. In Ref.~[\onlinecite{Cleland&GellerPRL04}] we presented resonance data down to $4.2 \, {\rm K}$ for a $1.8 \, {\rm GHz}$ AlN resonator. The observed low-temperature $Q$ of 3500 corresponds to an energy lifetime $\tau$ of more than $300 \, {\rm ns}$, already sufficient for most of the operations described below. This is to be contrasted with the previous state-of-the-art $1 \, {\rm GHz}$ SiC cantilever beam resonator demonstrated in 2003,\cite{HuangNat03} which has a $Q$ nearly an order of magnitude smaller at the same temperature. The unprecedented performance of our resonator is a consequence of the use of AlN, which is an intrinsically high $Q$ material, \cite{ClelandAPL01} and the use of the dilatational vibrational mode. 

The dilatational mode of interest is an approximately uniform oscillation of the thickness of the disk, which produces a nearly uniform electric field in a direction perpendicular to the disk. For a disk with large aspect ratio $R/b$, the dilatational mode frequency is
\begin{equation} 
\omega_0 \equiv \pi v/b,
\label{omega0 definition}
\end{equation}
where $v$ is a piezoelectrically enhanced sound speed to be defined below. Although the dilatational mode is not necessarily the fundamental mode of the resonator, we can couple to it by frequency selection, carefully avoiding the other low-frequency modes. The frequency in Eq.~(\ref{omega0 definition}) is that of the fundamental vibrational mode of a one-dimensional elastic string with free ends. For simplicity, we will assume that the dilatational mode frequency given by Eq.~(\ref{omega0 definition}) holds even if the aspect ratio $R/b$ is not large.

Quantum mechanically, each vibrational mode $n$ of such a resonator, having angular frequency $\omega_n$, is equivalent to a harmonic oscillator with energy level spacing $\hbar \omega_n$. For sufficiently high frequency and low temperature, the mode can be cooled to its quantum ground state: For example, if $ \omega_0/2 \pi = 15 \, {\rm GHz}$, then $\hbar \omega_0  /k_{\rm B}$ is about $720 \, {\rm mK}$. If cooled on a dilution refrigerator to $100 \, {\rm mK}$, the probability 
\begin{equation}
p_1 = 2 \,  {\rm sinh}({\textstyle{\hbar \omega_0 \over 2 k_{\rm B} T}}) \, e^{-3\hbar \omega_0/2 k_{\rm B} T}  
\end{equation} 
of thermally occupying the first excited (one-phonon) state, thereby producing a mixed state instead of the desired pure phonon ground state, is smaller than $10^{-3}$. The mean number $n_{\rm B}(\hbar \omega_0)$ of phonons present in the dilatational mode at $100 \, {\rm mK}$, or ``excitation level'' of the corresponding harmonic oscillator, is also less than $10^{-3}.$ Here $n_{\rm B}(\epsilon)$ is the Bose distribution function.

In the simulations below we will assume a nanomechanical disk resonator with the parameters given in Table \ref{resonator table}. The thickness $b$ is chosen to give a dilatational mode frequency $\omega_0/2 \pi $ of $15 \, {\rm GHz}$. This frequency is convenient for simulation because, when coupled to a JJ with parameters corresponding to that of Ref.~[\onlinecite{MartinisPRL02}], the bias current 
\begin{equation}
s^* \equiv \sqrt{1-(\omega_0/\omega_{{\rm p}0})^4}
\end{equation}
required to tune the qubit level spacing $\Delta E$ into resonance with $\hbar \omega_0$ is small enough so that the JJ eigenfunctions can be taken to be harmonic oscillator states. The resonator radius $R$ listed in Table \ref{resonator table} is chosen to make the junction-resonator interaction strength $g$, to be defined below, 1\% of $\hbar \omega_0 $, although we will also briefly consider larger resonators with larger interaction strengths. As stated above, the parameters listed in Table \ref{resonator table} assume that Eq.~(\ref{omega0 definition}) is valid. The AlN physical constants were obtained from the review by Ambacher.\cite{AmbacherJPD98}

\begin{table}
\caption{\label{resonator table}
Parameters characterizing the piezoelectric resonator simulated in this paper.}
\begin{ruledtabular}
\begin{tabular}{|c|c|}
piezoelectric material  & AlN \\ \hline
mass density $\rho$  & $3.26 \, {\rm g \, cm^{-3}}$ \\ \hline
dielectric constant $\epsilon_{33}/\epsilon_0$  & $10.7$  \\ \hline
elastic stiffness $c_{33}$  & $395 \, {\rm G Pa}$ \\ \hline
piezoelectric modulus $e_{33}$  & $1.46 \, {\rm C \, m^{-2}}$ \\ \hline
piezoelectric efficiency $\gamma \equiv e^2_{33}/ \epsilon_{33} c_{33}$  & $0.057$  \\ \hline
enhanced stiffness ${\tilde c}_{33} \equiv (1 + \gamma) c_{33} $  & $418 \, {\rm G Pa}$ \\ \hline
sound velocity $v \equiv \sqrt{{\tilde c}_{33}/\rho}$  & $11.3 \, {\rm km \, s^{-1}}$  \\ \hline
disk radius $R$  & $ 0.230 \, {\rm \mu m}$ \\ \hline
disk thickness $b$  & $377 \, {\rm nm} $ \\ \hline
dilatational frequency $\omega_0/2 \pi$  & $15 \, {\rm GHz} $ \\ \hline
frequency in Kelvin $\hbar \omega_0/ k_{\rm B}$  & $720 \, {\rm mK} $ \\ \hline
resonator capacitance $C_{\rm res}$ & $0.042 \, {\rm fF}$ \\ 
\end{tabular}
\end{ruledtabular}
\end{table}

We turn now to a calculation of the dilatational mode of the piezoelectric disk, assuming $b \ll R$. The disk lies in the $xy$ plane. In the $b \ll R$ limit, the elastic displacement field ${\bf u}({\bf r},t)$ for the dilatational mode is directed in the $z$ direction, and the $z$ component is itself only dependent on $z$ and $t$. Edge effects are assumed to be negligible. The vibrational dynamics for this mode and its harmonics is therefore effectively one-dimensional.

Let $u$ denote the $z$ component of the displacement field. To construct the equation of motion for $u(z,t)$, we write the basic electromechanical equations of piezoelectric media\cite{Auld1} in the form
\begin{equation}
E_z = {1 \over \epsilon_{33}} D_z - h_{33} \, \partial_z u,
\label{E equation}
\end{equation}
and
\begin{equation}
T_{zz} = - h_{33} \, D_z + {\tilde c}_{33} \, \partial_z u.
\label{T equation}
\end{equation}
Here $E_z$ and $D_z$ are the $z$ components of  the electric ${\bf E}$ and ${\bf D}$ fields, and $T_{ij}$ is the stress tensor. $\epsilon_{33}$ is the relevant element of the static dielectric tensor,  and $h_{33} \equiv e_{33}/\epsilon_{33}$, with $e_{33}$ the piezoelectric modulus. Finally, ${\tilde c}_{33} \equiv (1 + \gamma) c_{33}$ is a piezoelectrically enhanced elastic modulus, with $c_{33}$ denoting the appropriate element of the elastic tensor, and 
\begin{equation}
\gamma \equiv {e^2_{33} \over \epsilon_{33} c_{33}}
\label{piezoelectric efficiency definition}
\end{equation}
is a dimensionless quantity called the piezoelectric efficiency. The values of these material parameters for the case of AlN are summarized in Table \ref{resonator table}. Eq.~(\ref{E equation}) determines the relation between the electric field and strain inside the resonator, and Eq.~(\ref{T equation}) determines the stress-strain relationship, as modified by the electric field.

Electrically, the boundary conditions are that there is a charge per unit area $\sigma$ on the top electrode of a parallel-plate capacitor enclosing the resonator, and $-\sigma$ on the lower electrode. Then, in the interior of the piezoelectric, $D_z$ is uniform, with the value
\begin{equation}
D_z = -\sigma.
\end{equation}
Mechanically, the faces of the resonator are assumed to be stress free. We note from  Eq.~(\ref{T equation}) that when $\sigma \neq 0$, this stress-free condition requires a fixed {\it strain} of $ - h_{33} \sigma / {\tilde c}_{33}$ on the upper and lower surfaces of the disk. Note that these boundary conditions are generally time-dependent, because $\sigma$ usually is.

The resonator has thickness $b$ and occupies the region $0 <  z < b$. From the mechanical equation of motion $\rho \partial_t^2 u_i= - \partial_j T_{ij}$ we obtain
\begin{equation}
(\partial_t^2 - v^2 \partial_z^2)u = 0, \ \ \ {\rm with} \ \ \  v \equiv \sqrt{{\tilde c}_{33}/\rho}.
\label{wave equation}
\end{equation}
The sound velocity in the $z$ direction is slightly enhanced because of the piezoelectric effect. The most general solution of Eq.~(\ref{wave equation}), satisfying the required boundary conditions, is
\begin{equation}
u(z,t) = - {h_{33} \sigma(t) \over {\tilde c}_{33} } \, z + {\rm Re} \, \sum_{n=0}^\infty A_n \cos(k_n z) \, e^{-i v k_n t} \! ,
\label{general solution}
\end{equation}
where
\begin{equation}
k_n \equiv n \pi /b.
\label{k definition}
\end{equation} 
Here we have assumed that $\sigma$ is quasi-stationary, so that $\partial_t^2 \sigma$ is negligible. The first term in Eq.~(\ref{general solution}) describes a background strain caused the electric field in the capacitor, present in the classical limit even at zero temperature, while the second term describes harmonic fluctuations about that strain. The $n=0$ mode is a center-of-mass translation. The $n=1$ mode is the fundamental thickness-oscillation mode of interest here; it has an angular frequency given by Eq.~(\ref{omega0 definition}).

\subsection{Model Hamiltonian\label{model hamiltonian section}}

Next we derive a model Hamiltonian for a single current-biased JJ coupled to the dilatational mode of a piezoelectric nanomechanical disk resonator. The layout is similar to that illustrated in Fig.~\ref{fig.twoqubit}, except that there is only one junction, and the gate electrode is not split.  Extension to multiple junctions and resonators will be carried out in Sec.~\ref{network section}. As before, we will assume that the junction and resonator states are long lived, and any effects of decoherence are neglected. We will proceed by returning to the semiclassical description of the JJ reviewed in Sec.~\ref{semiclassical section}, including the resonator in the equivalent circuit, and then requantizing the coupled system.

Our first objective is to derive an equation for $I_{\rm res}$, the resonator's contribution to the electrical current seen by the JJ. $I_{\rm res}$ is equal to ${\dot q}$, where $q$ is the charge on the resonator's top (ungrounded) electrode produced by voltage fluctuations across and strain fluctuations inside the resonator.
Integrating Eq.~(\ref{E equation}) gives the voltage
\begin{equation}
V = - \int_0^b dz \, E_z = {\sigma b \over \epsilon_{33}} + h_{33} b U,
\label{V}
\end{equation}
across the resonator and JJ, in terms of the charge on the electrodes and the spatially averaged strain
\begin{equation}
U(t) \equiv {u(b,t) - u(0,t) \over b}
\label{average strain definition}
\end{equation}
in the resonator. Eq.~(\ref{V}) can then be written in terms of the total charge $q \equiv \sigma \pi R^2$ on the upper plate as
\begin{equation}
q = C_{\rm res} \big( V - b \, h_{33} \, U \big),
\end{equation}
where $C_{\rm res} \equiv \epsilon_{33} \pi R^2/b$ is the geometric capacitance of the resonator. 

The resonator therefore produces a current equal to
\begin{equation}
I_{\rm res} =  C_{\rm res} \big( {\dot V} - b \, h_{33} \, {\dot U} \big).
\label{Ires}
\end{equation}
The first term in Eq.~(\ref{Ires}) describes a purely capacitive effect, which would be present even in the absence of the piezoelectric disk between the electrodes. We will find that this term simply adds the capacitance of the resonator in parallel with the junction capacitance $C$, thereby reducing the junction's charging energy. The second term is a consequence of piezoelectricity, and will be shown to have two effects: coupling the JJ to resonator phonons and renormalizing $C_{\rm res}$.

It will be convenient to write Eq.~(\ref{general solution}) as
\begin{equation}
u(z,t) = - {h_{33} \sigma(t) \over {\tilde c}_{33} } \, z + \delta u(z,t),
\label{general solution form}
\end{equation}
where
\begin{equation}
\delta u(z,t) \equiv  {\rm Re} \, \sum_{n=0}^\infty A_n \cos(n \pi z/b) \, e^{-i v k_n t} 
\label{fluctuation definition}
\end{equation}
is the harmonic fluctuation contribution. After quantization, this latter part of the displacement field will come from phonons. The average strain can be similarly expanded as
\begin{equation}
U(t) = - {h_{33} \sigma(t) \over {\tilde c}_{33} } + \delta U(t),
\label{strain expansion}
\end{equation}
where
\begin{equation}
\delta U(t) \equiv {\delta u(b,t) - \delta u(0,t) \over b}.
\end{equation}

Now, the time derivative of the first term in Eq.~(\ref{strain expansion}) is itself proportional to $I_{\rm res}$, so Eq.~(\ref{Ires}) can be equivalently written as
\begin{equation}
I_{\rm res} =   {\tilde C}_{\rm res} \big( {\dot V} - b \, h_{33} \, {\delta \dot U} \big),
\label{new Ires}
\end{equation}
where 
\begin{equation}
{\tilde C}_{\rm res} \equiv {C_{\rm res} \over 1-\gamma - \gamma^2}
\end{equation}
is a piezoelectrically enhanced resonator capacitance, and $\gamma$ is the piezoelectric efficiency defined in Eq.~(\ref{piezoelectric efficiency definition}). In contrast with that of Eq.~(\ref{Ires}), the second term in Eq.~(\ref{new Ires}) describes a pure coupling to resonator phonons.

Returning to the inset of Fig.~\ref{fig.washboard}, we replace $I_{\rm b}$ with $I_{\rm b} + I_{\rm res}.$ In our coupled junction-resonator system, $I_{\rm b}$ then refers to the bias current coming from the external circuitry alone, which may have both DC and RF components (see Sec.~\ref{preparation and readout section}). The semiclassical equation of motion replacing Eq.~(\ref{equation of motion}) is now that of a particle with a modified mass moving in a potential $U + \delta H_{\rm cl}$, where
\begin{equation}
\delta H_{\rm cl} \equiv {\hbar C_{\rm res} b \, h_{33} \, \delta {\dot U} \over 2 e (1-\gamma - \gamma^2)} \, \delta .
\label{classical interaction}
\end{equation}
The classical junction-resonator interaction Hamiltonian $\delta H_{\rm cl}$ is evidently linear in the phase difference $\delta$. The effective mass $M$ of the particle is given by Eq.~(\ref{effective mass}), with $E_{\rm c}$ now reduced to $2e^2/(C+{\tilde C}_{\rm res}).$

Quantization of the $\delta$ variable proceeds as in Sec.~\ref{quantization section}. The quantization of the resonator dynamics is carried out in Appendix \ref{resonator appendix}. The resonator Hamiltonian (dropping an irrelevant additive constant) is
\begin{equation}
H_{\rm res} = \hbar \omega_0 a^\dagger a,
\label{resonator hamiltonian}
\end{equation}
where $a^\dagger$ and $a$ are bosonic creation and annihilation operators for dilatational phonons. The junction-resonator interaction Hamiltonian is found to be
\begin{equation}
\delta H \equiv  -i g (a - a^\dagger)\delta,
\label{interaction hamiltonian}
\end{equation}
where
\begin{equation}
g \equiv \frac{\hbar^{3/2} \, e_{33} \, {\tilde C}_{\rm res}  \, \sqrt{\omega_0} } {e \, \epsilon_{33} \sqrt{\rho \pi R^2 b \,}} 
\label{coupling strength formula}
\end{equation}
is a real-valued coupling constant with dimensions of energy. We note that $g$ depends only on the properties of the resonator and is independent of the parameters characterizing the Josephson junction. The value of $g$ quoted in Eq.~(\ref{coupling strength formula}) applies to a fully gated resonator coupled to a single JJ; for a JJ connected to one half of a split-gate resonator, such as in Fig.~\ref{fig.twoqubit}, the relevant interaction strength is $g/2$.

\begin{table}
\caption{\label{junction-resonator table}
Parameters for a single JJ coupled to the resonator of Table \ref{resonator table}. The junction parameters correspond to that investigated Ref.~[\onlinecite{MartinisPRL02}].}
\begin{ruledtabular}
\begin{tabular}{|c|c|}
critical current $I_0$  & $ 21 \, \mu {\rm A}$ \\ \hline
Josephson energy $E_{\rm J}$  & $43.05 \, {\rm meV}$ \\ \hline
junction capacitance $C$  & $6 \, {\rm pF}$ \\ \hline
charging energy $E_{\rm c} $  & $53.33 \, {\rm neV}$ \\ \hline
zero-bias plasma frequency $\omega_{{\rm p}0}/2 \pi $  & $16.4 \, {\rm GHz}$ \\ \hline
resonant bias current $s^*$  & 0.545 \\ \hline
junction-resonator interaction strength $g$ & $0.620 \, {\rm \mu eV}$ \\ \hline
resonant vacuum Rabi frequency $\Omega(0)/2 \pi$ & $8.79 \, {\rm MHz}$ \\ \hline
resonant Rabi period $2 \pi/ \Omega(0)$ & $113.7 \, {\rm ns}$ \\ 
\end{tabular}
\end{ruledtabular}
\end{table}

For a fixed disk thickness $b$, chosen to determine $\omega_0$, the interaction strength varies linearly with disk radius $R$. Using the parameters summarized in Table \ref{resonator table} for a $15 \, {\rm GHz}$ AlN resonator, we obtain
\begin{equation}
g = 2.70 \, {\rm \mu eV} \times R[{\rm \mu m}],
\label{numerical coupling strength formula}
\end{equation}
where $R[{\rm \mu m}]$ is the resonator radius in ${\mu m}.$ In the simplest qubit storage simulations carried out below, we choose $R$ to be $0.230 \, {\rm \mu m}$, in which case the interaction strength is $0.620 \, {\rm \mu eV}.$ In Table \ref{junction-resonator table} we summarize this and other parameters associated with the most basic coupled JJ-resonator system.

The complete Hamiltonian of the system is
\begin{equation}
H =  H_0 + \delta H, \ \ \ {\rm with} \ \ \ H_0 \equiv H_{\rm J}+H_{\rm res}.
\label{full hamiltonian}
\end{equation}
The junction Hamiltonian $H_{\rm J}$ depends on $s$, and when $s$ is time-dependent, $H_{\rm J}$ is also time-dependent. We shall address this issue below in Sec.~\ref{dynamics section}. Assuming $s$ is constant, the stationary states of $H_0$ may be written as
\begin{equation}
|mn\rangle \equiv |m\rangle_{\rm J} \otimes |n\rangle_{\rm res},
\end{equation}
where $m=0,1,2,\dots$ labels the junction state and $n=0,1,2,\dots$ is the phonon occupation number of the resonator. The eigenvalues of $H_0$ are
\begin{equation}
E_{mn} =  \epsilon_m  +  \hbar \omega_0 \, n.
\end{equation}
The $|mn\rangle$ and $E_{mn}$ of course depend on $s$. We will refer to the lowest two eigenstates of $H_0$ as the phase qubit, and to $\Delta E$ [defined in Eq.~(\ref{level spacing})] as the qubit level spacing, even if there are more than two quasibound levels in the junction.

For many applications it is convenient to write the JJ Hamiltonian of Eq.~(\ref{junction hamiltonian}) in second-quantized form, as
\begin{equation}
H_J = \sum_m \epsilon_m c_m^\dagger c_m. 
\label{second quantized junction hamiltonian}
\end{equation}
Here $c^\dagger_m$ and $c_m$ are creation and annihilation operators for the junction states, which can be taken to be either fermionic or bosonic because there is only one ``particle'' in the washboard potential. In this same notation, the interaction Hamiltonian becomes
\begin{equation}
\delta H = - i g \sum_{mm'} x_{mm'} \,  (a - a^\dagger) \, c_m^\dagger c_{m'},
\label{second quantized interaction hamiltonian}
\end{equation}
where the $x_{mm'}$ are dipole-moment matrix elements defined in Eq.~(\ref{dipole moment definition}). 

%\begin{figure}
%\includegraphics[width=6.0cm]{fig5.jpg}
%\caption{\label{fig.fourqubits} Four current-biased JJs coupled to a nanoelectromechanical resonator. Each junction is connected to a metallic plate on the surface of the resonator that covers about one quarter of the surface. Because we make use of the fundamental dilatational mode, which is spatially uniform in the plane of the resonator, the qubits are all equally well coupled to that mode.}
%\end{figure}

An important simplification occurs when only the qubit states $m=0,1$ are included in the JJ. In this case the complete Hamiltonian can be written as 
\begin{equation}
H =  \begin{pmatrix} \epsilon_0 & 0 \\ 0 & \epsilon_1 \end{pmatrix} + \ \hbar \omega_0 a^\dagger a  -  i g (a - a^\dagger) \! \! 
\begin{pmatrix} x_{00} & x_{01} \\ x_{10} & x_{11} \end{pmatrix} \! \! ,
\label{matrix H}
\end{equation}
with the JJ operators written as matrices in the $\lbrace |0\rangle, |1\rangle \rbrace$ basis. Recall that the diagonal dipole moments $x_{mm}$ do not generally vanish in the current-biased JJ, but for $s$ not too close to unity are approximately equal to ${\rm arcsin} \, s.$ In the approximation $x_{00} = x_{11}$, and dropping an additive constant, we can succinctly write (\ref{matrix H}) in terms of the Pauli matrices as
\begin{equation}
H =  - {\textstyle{ \Delta E \over 2}} \sigma_z+  \hbar \omega_0 a^\dagger a  -  i g (a - a^\dagger) [x_{00} \sigma_0 + x_{01} \sigma_x ],
\label{pauli H}
\end{equation}
where $\sigma_0$ is the identity matrix. Note, however, that $x_{00} \neq x_{11}$ when $s$ is very close to 1. Finally, when both $\Delta E \approx \hbar \omega_0$ and $g \ll \Delta E$, the commonly used rotating-wave approximation of quantum optics becomes valid. Applied to the form (\ref{matrix H}) or (\ref{pauli H}), the Hamiltonian simplifies to 
\begin{equation}
H_{\rm JC} \equiv  - {\textstyle{ \Delta E \over 2}} \sigma_z+  \hbar \omega_0 a^\dagger a -  i g x_{01} \big( a \sigma_{-} - a^\dagger \sigma_{+} \big),
\label{JC model}
\end{equation}
where $\sigma_{\pm} \equiv (\sigma_x \pm i \sigma_y)/2$. $H_{\rm JC}$ is the Jaynes-Cummings model of quantum optics (written is a basis that is different from that conventionally used there).

The Hamiltonian in Eq.~(\ref{full hamiltonian}) is equivalent to that of a few-level atom in an electromagnetic cavity. The JJs are analogous to the atoms. The cavity photons here are dilatational-mode phonons, which interact electrically with the junctions via the piezoelectric effect. Coupling several junctions to a nanomechanical resonator, as illustrated in Fig.~\ref{fig.fourqubits}, then makes the system analogous to several atoms in an electromagnetic cavity, except that here the atomic level spacing and electron-photon interaction strength are all externally controllable.

\subsection{Quantum dynamics in the instantaneous basis\label{dynamics section}}

As discussed above, the Hamiltonian $H_{\rm J}$ for the JJ depends on the dimensionless bias current $s$, and is therefore usually time-dependent. It will be useful to expand the state of the coupled system in a basis of normalized instantaneous eigenstates $|mn\rangle_{\! s}$ of $H_0$, defined by
\begin{equation}
H_0(s) \big|mn\big\rangle_{\! s} = E_{mn}(s) \big|mn\big\rangle_{\! s}, \ \ \ {\rm with} \ \ \  s=s(t).
\end{equation}
We assume that at time $t=t_0$ the bias current is constant and that the system is prepared in a pure state. For $t>t_0$ we write the wave function, suppressing the time-dependence of $s(t),$ as
\begin{equation}
\big|\psi(t)\big\rangle = \sum_{mn} c_{mn}(t) \, e^{-(i/\hbar) \! \int_{t_0}^t dt' E_{mn}(s)} \big|mn\big\rangle_{\! s}.
\label{general wave function expansion}
\end{equation}
The probability amplitudes in the instantaneous interaction representation satisfy
\begin{eqnarray}
i \hbar {\dot c}_{mn} &=& \sum_{m'n'} \langle mn | \delta H - i \hbar \partial_t |m'n'\rangle_{\! s}  \nonumber \\
&\times& e^{(i/\hbar) \! \int_{t_0}^t \! dt' [E_{mn}(s) - E_{m'n'}(s)]} \ c_{m'n'}.
\label{schrodinger equation}
\end{eqnarray}
Off-diagonal matrix elements of the quantity
\begin{equation}
\langle mn | {\textstyle{\partial \over \partial t}} |m'n'\rangle_{\! s} = \langle mn | {\textstyle{\partial \over \partial s}} |m'n'\rangle_{\! s} \, {\dot s}
\end{equation}
determine transitions between the instantaneous eigenstates caused by nonadiabatic variation of $s$; the diagonal elements determine the Berry connection of adiabatic perturbation theory. In the small $s$, quadratic-potential limit, the low-lying JJ eigenstates are well approximated by the harmonic oscillator eigenfunctions given in Eq.~(\ref{oscillator basis functions}). In this case it can be shown that
\begin{eqnarray}
\langle mn | {\textstyle{\partial \over \partial s}} |m'n'\rangle_{\! s} &=& {\textstyle{1 \over \ell_s \sqrt{1-s^2}}} \big( {\textstyle\sqrt{m' + 1\over 2}} \, \delta_{m,m'+1}  \nonumber \\
&-& {\textstyle\sqrt{m' \over 2}} \, \delta_{m,m'-1} \big) \delta_{nn'} \nonumber \\
&+& {\textstyle{1 \over \ell_s}} {\textstyle{d \ell_s \over ds}} \big( {\textstyle{\sqrt{(m' + 1)(m'+2)}\over 2}} \, \delta_{m,m'+2}  \nonumber \\
&-&  {\textstyle{\sqrt{m'(m'-1)}\over 2}} \, \delta_{m,m'-2} \big) \delta_{nn'},
\label{nonadiabatic elements}
\end{eqnarray}
where
\begin{equation}
{d \ell_s \over ds} = { s \ell_s \over 4(1-s^2)}.
\end{equation}
There are no diagonal (Berry connection) terms in this limit. The terms in Eq.~(\ref{nonadiabatic elements}) proportional to $d\ell_s /ds$ result from the change of curvature at the minimum of the {\it anharmonic} potential $U(\delta)$ with changing $s$.

\section{QUBIT STORAGE AND TRANSFER\label{qubit storage and transfer section}}

We now turn to a discussion of some single-qubit operations made possible by the nanomechanical resonator. In particular, we show that any phase qubit state
\begin{equation}
|\psi_{\rm J} \rangle = \alpha |0\rangle_{\rm J} + \beta |1\rangle_{\rm J} \ \ \ \ \ {\rm with} \ \ \ \ \ |\alpha|^2 + |\beta|^2 = 1
\label{junction state}
\end{equation}
produced in the current-biased JJ can be written to and coherently stored in the phonon-number states of the resonator, as
\begin{equation}
|\psi_{\rm res} \rangle = \alpha |0\rangle_{\rm res} + \beta |1\rangle_{\rm res},
\label{resonator state}
\end{equation}
yielding a quantum memory element. In Eq.~(\ref{junction state}), the states $|0\rangle_{\rm J}$ and $|1\rangle_{\rm J}$ are the lowest two junction eigenstates shown in 
Fig.~\ref{fig.currentbiasjj}, whereas in Eq.~(\ref{resonator state}) they denote the vacuum and one-phonon states of the resonator's dilatational mode. Later, the qubit state can be reversibly retrieved or transferred to a  second Josephson junction.

We will examine qubit storage and transfer in two stages: First we will develop a simple analytic theory based on the adiabatic approximation combined with the rotating-wave 
approximation (RWA) of quantum optics.\cite{Scully} The adiabatic approximation assumes that the bias current $s$ changes slowly on the frequency scale $\Delta E/\hbar$, 
a requirement that (although not always desirable) can be easily satisfied in practice. The RWA for a phase qubit is valid when two conditions are met: 
\begin{itemize}
\item[(i)]  $\Delta E$ and $\hbar \omega_0$ are close to each other on the scale of the resonator's energy width $\hbar \omega_0/Q$. Here $Q$ is the resonator's dilatational-mode quality factor. Transitions to higher levels $|m\rangle_{\rm J}$ with $m>1$ are far off resonance on this same scale. 
\item[(ii)]  The interaction strength $g$ is small compared with $\Delta E$ (or $\hbar \omega_0$).
\end{itemize} 
We will then supplement the analytic theory with numerical simulations based on the full Hamiltonian of Eq.~(\ref{full hamiltonian}), using realistic values of all parameters involved.

\subsection{RWA analysis\label{RWA section}}

To understand qubit storage, consider a single junction coupled to a nanomechanical resonator as described by the Hamiltonian of Eq.~(\ref{full hamiltonian}), and expand the 
wave function for the combined system as in Eq.~(\ref{general wave function expansion}). The probability amplitudes $c_{mn}(t)$ in the instantaneous interaction representation then satisfy Eq.~(\ref{schrodinger equation}).

We start at some time $t_0 < 0$ with the JJ prepared in the state (\ref{junction state}) and the resonator in its ground state,
\begin{equation}
|\psi(t_0)\rangle = \big(\alpha |0\rangle_{\rm J} + \beta |1\rangle_{\rm J} \big) \! \otimes \! |0\rangle_{\rm res} = \alpha |00\rangle + \beta |10\rangle.
\label{storage initial state}
\end{equation}
We assume that the qubit and resonator are detuned and that $g \ll \Delta E$. Then the $|mn\rangle$ in (\ref{storage initial state}) are close to eigenstates, and the $c_{mn}$ remain approximately constant. The qubit level spacing $\Delta E$ is now adiabatically changed to the resonant value, reaching $\hbar \omega_0$ at time $t=0.$ Then at $t=0$ we have, approximately,
\begin{equation}
c_{mn}(0) = \big( \alpha \, \delta_{m0} + \beta \, \delta_{m1} \big) \delta_{n0}. 
\label{storage initial condition}
\end{equation}
As we shall discuss below, the first nonadiabatic corrections to Eq.~(\ref{storage initial condition}) principally affect the phases of the $|00\rangle$ and $|10\rangle$ components of the wave function, that is, the phases of $\alpha$ and $\beta$. The wave function at $t=0$ is therefore
\begin{eqnarray}
\big|\psi(0)\big\rangle &\approx&  \alpha \, e^{-(i/\hbar)\int_{t_0}^0 \! dt E_{00}[s(t)]} \, \big|00\big\rangle \nonumber \\ 
&+& \beta \, e^{-(i/\hbar)\int_{t_0}^0 \! dt E_{10}[s(t)]} \, \big|10\big\rangle.
\label{t=0 state}
\end{eqnarray}

Next we invoke the RWA, which allows us to write Eq.~(\ref{schrodinger equation}) as
\begin{eqnarray}
{\dot c}_{0n} &=& \frac{g}{\hbar} \, \sqrt{n} \  x_{01} \,e^{i \omega_{\rm d} t} \, c_{1,n-1} \nonumber \\
{\dot c}_{1n} &=&  -\frac{g}{\hbar} \, \sqrt{n+1} \  x_{01} \, e^{-i \omega_{\rm d} t} \, c_{0,n+1}. 
\label{RWA equations}
\end{eqnarray}
We have also assumed that all dissipation and decoherence mechanisms are negligible over experimental timescales. Furthermore, as discussed in Sec.~\ref{quantization section}, we can take $x_{01}$ to be positive with no loss of generality. Using Eqs.~(\ref{storage initial condition}) and (\ref{RWA equations}), we then obtain, by Laplace transformation,
\begin{eqnarray}
c_{00}(t) &=& \alpha \nonumber \\
c_{01}(t) &=& \beta \, {\Omega(0) \over \Omega} \, \sin({\textstyle{\Omega t \over 2}}) \, e^{i \omega_{\rm d} t/2} \nonumber \\
c_{10}(t) &=& \beta \bigg[ \cos({\textstyle{\Omega t \over 2}})+ i {\omega_{\rm d} \over \Omega} \sin({\textstyle{\Omega t \over 2}}) \bigg]  e^{-i \omega_{\rm d} t/2} \nonumber \\
c_{11}(t) &=& 0,
\end{eqnarray}
and all $c_{mn}(t)$ with $n>1$ equal to zero. Here
\begin{equation}
\Omega(\omega_{\rm d}) \equiv \sqrt{ [\Omega(0)]^2 + \omega_{\rm d}^2} \ \ \ {\rm with} \ \ \ \Omega(0) \equiv {2 g x_{01} \over \hbar},
\label{vacuum Rabi frequency definition}
\end{equation}
is the vacuum Rabi frequency, and $\omega_{\rm d} \equiv \omega_0 - \Delta E /\hbar$ is the resonator-qubit detuning. $\Omega(0)$ is the Rabi frequency on resonance. Probability amplitudes at selected times are summarized in Table~\ref{amplitude table}. The wave function at later times $t > 0$, when the system is on resonance, is therefore
\begin{eqnarray}
{\hskip -0.4in} \big|\psi(t)\big\rangle &\approx& e^{- i E_{00}[s^*] t/\hbar} \bigg( \alpha e^{-(i/\hbar) \! \int_{t_0}^0 \! \!  dt E_{00}[s(t)]} \big|00\big\rangle \nonumber \\
&+& \beta \sin({\textstyle{\Omega t \over 2}})  e^{-(i/\hbar) \! \int_{t_0}^0 \! \! dt E_{01}[s(t)]} \, e^{- i \omega_0 t} \,  \big|01\big\rangle \nonumber \\
&+& \beta \cos({\textstyle{\Omega t \over 2}}) e^{-(i/\hbar) \! \int_{t_0}^0 \! \! dt E_{10}[s(t)]} \, e^{- i \omega_0 t} \, \big|10\big\rangle \bigg),
\label{state during pulse}
\end{eqnarray}
where $s^*$ is the resonant value of the dimensionless bias current. We emphasize that the result in Eq.~(\ref{state during pulse}) is only approximate.

\begin{table}
\caption{\label{amplitude table}
Probability amplitudes $c_{mn}(t)$ for phase-qubit coupled to nanomechanical resonator, at time zero when they are brought to perfect resonance, as well as one quarter, one half, and three quarters of a vacuum Rabi-oscillation period later.}
\begin{ruledtabular}
\begin{tabular}{|c|cccc|}
amplitude & $t=0$ & $t =\pi/2\Omega$ & $t  =\pi/\Omega$ & $t =3\pi/2\Omega$ \\ \hline
$c_{00}$ & $\alpha$ & $\alpha$ & $\alpha$ & $\alpha$  \\
$c_{01}$ & $0$ & $\beta/\sqrt{2}$ & $\beta$ & $\beta/\sqrt{2}$ \\
$c_{10}$ & $\beta$ & $\beta/\sqrt{2}$ & $0$ & $-\beta/\sqrt{2}$ \\
$c_{11}$ & $0$ & $0$ & $0$ & $0$
\end{tabular}
\end{ruledtabular}
\end{table}

After a pulse duration $\Delta t$, the JJ is again detuned from the resonator. The final wave functions, in the instantaneous interaction representation, for several important choices of $\Delta t$ is summarized in Table \ref{pulse table}. In this representation the phase factors $\exp[-(i/\hbar) \! \int \! dt \, E_{mn}(s)]$ appearing in Eq.~(\ref{general wave function expansion}) are suppressed.

When $\Omega \, \Delta t= \pi/2$, the system is held in resonance for one quarter of the vacuum Rabi period, and the final state is entangled. In particular, when the qubit begins completely in the excited state, $\alpha = 0$ and $\beta=1$, the resulting state is the maximally entangled Bell state
\begin{equation}
{ |01\rangle_{\rm J} + |10\rangle_{\rm J} \over \sqrt{2}} \otimes |0\rangle_{\rm res} \, ,
\end{equation}
with the resonator returned to the ground state. Similarly, after three quarters of a Rabi period, the final state is $ 2^{-{1 \over 2}}(|01\rangle_{\rm J} - |10\rangle_{\rm J}) \otimes |0\rangle_{\rm res}.$

After half a Rabi period, or $\Omega \, \Delta t= \pi$, the phase qubit and resonator states are evidently swapped. The cavity-QED analog of this operation has been demonstrated experimentally in Ref.~[\onlinecite{MaitrePRL97}]. This operation is extremely useful in our architecture. In addition to allowing the resonator to be used as a quantum memory element, it can be used as a quantum bus to transfer a qubit state from one JJ to another attached to the same resonator. 

To retrieve a state that has been stored in the resonator, the junction is again tuned to $\hbar \omega_0$, except this time for one and a half Rabi periods, or $\Omega \, \Delta t = 3 \pi.$ This longer pulse length requirement follows from the RWA equations (\ref{RWA equations}). An exception is the special case where the stored state is a $|0\rangle$ or $|1\rangle$, and one does not care about the overall phase of the final result, as in our Ref.~[\onlinecite{Cleland&GellerPRL04}]. We have extensively simulated the use of the resonator as a quantum memory element, and its speed and fidelity as a function of coupling strength and position of the stored state on the Bloch sphere. With dimensionless interaction strengths $g /\hbar \omega_0$ around a few percent, the RWA is quite reliable, and the main source of error comes from nonadiabatic effects during the ramping of $s(t)$, which leads to errors in Eq.~(\ref{storage initial condition}). These are principally phase errors in $c_{00}$ and $c_{10}$, which results in a significant qubit-state dependence to the memory fidelity, with states closer to the equator of the Bloch sphere being stored less accurately. We shall return to these issues in future work.

To transfer a qubit state $\alpha |0\rangle_{{\rm J}1} + \beta |1\rangle_{{\rm J}1}$ from junction 1 to junction 2, the state is stored in the resonator's dilatational phonon number states as $\alpha |0\rangle_{\rm res} + \beta |1\rangle_{\rm res}$. After junction 1 is taken out of resonance, the bias on the junction 2 is varied to bring it into resonance with the resonator for one and a half Rabi periods ($\Omega \, \Delta t = 3 \pi),$ resulting in the creation of the state $\alpha |0\rangle_{{\rm J}2} + \beta |1\rangle_{{\rm J}2}$ in the second junction. (Again, the case where the transferred state is a $|0\rangle$ or $|1\rangle$ is exceptional, and a half of a Rabi period is sufficient.) The original qubit state is therefore transferred from one junction to another. It will be possible to verify experimentally that this has occurred by reading out the second junction at the end of the transfer operation.

\begin{table}
\caption{\label{pulse table}
Approximate final wave functions, in the instantaneous interaction representation, after the phase qubit and resonator have been in resonance for a time $\Delta t$.}
\begin{ruledtabular}
\begin{tabular}{|c|cc|}
$\Omega \, \Delta t$ & final state & operation \\ \hline
$ \pi/2 $ & $\alpha |00\rangle +\beta(|01\rangle+|10\rangle)/\sqrt{2}$ & entangle \\
$ \pi $ & $ |0\rangle_{\rm J} \! \otimes \! (\alpha |0\rangle_{\rm res} + \beta |1\rangle_{\rm res})$ & swap \\
$ 3\pi/2 $ & $\alpha |00\rangle +\beta(|01\rangle-|10\rangle)/\sqrt{2}$ & entangle 
\end{tabular}
\end{ruledtabular}
\end{table}

\subsection{Simulating storage and transfer\label{storage and transfer section}}

The analysis above, which is based on the adiabatic and rotating-wave approximations, implies that JJ states can be stored, transferred, and controllably entangled with perfect accuracy, and---with an appropriate choice of $g$---arbitrarily quickly. This is not the case: The actual fidelity is determined by the corrections to these approximations. In this section we shall study the storage and transfer fidelities by direct numerical integration of the time-dependent Schr\"odinger equation.

We begin by simulating the storage of a JJ state in the phonon-number states of a resonator. To do this we solve the time-dependent Schr\"odinger equation for the coupled 
junction-resonator system by numerically integrating the coupled equations (\ref{schrodinger equation}) for the case [see Eq.~(\ref{junction state})] 
\begin{equation}
\alpha = 0 \ \ \ \ {\rm and} \ \ \ \ \beta = 1.
\end{equation}
This corresponds to the phase qubit starting in the excited eigenstate $|1\rangle_{\! \rm J}$.  The resonator starts out in its ground state $|0\rangle_{\rm res}$. Our main result,
which is shown in Fig.~\ref{fig.storage}, will be discussed in detail below.

To ensure the reliability of the numerical results we employed a variety of ODE integrators, including both explicit and implicit algorithms, as well as exact diagonalization for cases with constant $s$. No significant differences were observed. The results presented were obtained with the 4th-order Runge-Kutta method with a time step of $1 \, {\rm fs}$, which guaranteed that probability was conserved for the duration of the calculation to better than 99.993\%. Josephson junction energy levels $\epsilon_m$ and dipole-moment matrix elements $x_{mm'}$ as a function of $s$ were calculated using the diagonalization method discussed in Sec.~\ref{quantization section}, and found to be extremely close to that of a harmonic oscillator in the range of bias currents employed here.

%\begin{figure}
%\includegraphics[width=.50\textwidth]{fig6.jpg}
%\caption[]{\label{fig.storage}Phase qubit storage. The solid descending curve is $|c_{10}(t)|^2$, the interaction-representation occupation probability of the $|10\rangle$ state, calculated numerically for the junction of Ref.~[\onlinecite{MartinisPRL02}] coupled to the $15 \, {\rm GHz}$ piezoelectric resonator described in Table \ref{resonator table}. The dashed curve is the same quantity calculated from the analytic RWA results of Sec.~\ref{RWA section}. The solid ascending curve is $|c_{01}(t)|^2$. The dotted curve shows the time dependence of the dimensionless bias current $s(t)$, which is varied to bring the phase qubit in resonance with the resonator after $5 \, {\rm ns}$.  The Rabi period on resonance, when $s=0.546$, is $113.69 \, {\rm ns}$. After the storage operation, $|c_{10}|^2 = 0.002$ and $|c_{01}|^2 = 0.987$. The inset shows an enlarged view of $|c_{10}(t)|^2$ during the ramping up of $s(t)$.}
%\end{figure}

We simulate a large area, current-biased JJ with parameters corresponding to that investigated in Ref.~[\onlinecite{MartinisPRL02}], namely $E_{\rm J} = 43.05 \, {\rm meV}$ and $E_{\rm c} = 53.33 \, {\rm neV}$. The zero-bias plasma frequency $\omega_{{\rm p}0}/2 \pi$ is therefore $16.4 \, {\rm GHz}$.  A $15 \, {\rm GHz}$ resonator will be in resonance with this junction when $s=0.545$, comfortably far from the regime near $s=1$ where bias-current fluctuations are most destructive. The nanomechanical resonator we simulate has the parameters listed Table \ref{resonator table}, which results in a junction-resonator interaction strength $g$ given in Table \ref{junction-resonator table}. The resonator thickness $d$ is determined by the desired $15 \, {\rm GHz}$ frequency of the thickness-oscillation mode, and the disk radius $R$ can be used to vary $g$ without appreciably affecting that frequency. As we noted in Eq.~(\ref{numerical coupling strength formula}), $g$ is linearly proportional to $R$ (in the large $R/d$ limit). We have used this tunability to ensure that the system is in the regime where the RWA analysis of Sec.~\ref{RWA section} is applicable. Below we will briefly examine results of simulations with larger values of $g$. There are more than 400 quasibound states $|m\rangle_{\rm J}$ in the junction when $s=0.545$. To the accuracy of the numerical results reported here, we find no sensitivity to the number of JJ states included in the calculations as long as at least 4 states are included. The resonator, of course, has an infinite number of phonon-number eigenstates $|n\rangle_{\rm res}$, and the results shown here have been calculated by including the 4 states lowest in energy, as increasing beyond this number led to no significant changes.

We turn now to a discussion of Fig.~\ref{fig.storage}. At time zero the current bias is $s=0.40$ and the wave function amplitudes are taken to be $c_{mn}(0) = \delta_{m1}  \delta_{n0}.$ The bias is held at $s=0.40$ for $5 \, {\rm ns}$. As shown in Fig.~\ref{fig.storage}, the occupation probability of the $|10\rangle$ state remains close to unity during this time interval. All other states remain essentially unoccupied. After $5 \, {\rm ns}$ the bias current is adiabatically changed to the resonant value of $s=0.545$. Our simulations show that the success of a qubit storage depends somewhat sensitively on the {\it shape} of the bias-current profile $s(t)$ in the transition region. In particular, we find that the time during which $s$ changes from the off-resonant value to the resonant one should be at least exponentially localized. The result presented in Fig.~\ref{fig.storage} was obtained using a trapezoidal profile with a cross-over time of $1 \, {\rm ns}$, which should be compared with the resonator and on-resonance qubit period of $0.1 \, {\rm ns}.$ Similar results were obtained using Gaussian profiles. The JJ level spacing is tuned to $\hbar \omega_0$ for half of a Rabi period $\pi/\Omega$. During this time interval the junction interacts strongly with the resonator, and energy is exchanged back and forth between the two systems. The JJ is then detuned from the resonator. Some of the final probability amplitudes are given in Table \ref{storage table}. For the small value of $g$ used here, chosen so that $g/\hbar \omega_0 = 0.01,$ the numerical results for the $|c_{mn}|^2$ are in excellent agreement with the RWA. However, the RWA prediction for the phases of the $c_{mn}$ are poor until one goes to even smaller values of $g$. In other words, the RWA is better at predicting the moduli of the $c_{mn}$ than their phases.

\begin{table}
\caption{\label{storage table}
Final state amplitudes $c_{mn}$ after qubit storage. System parameters are the same as in Fig.~\ref{fig.storage}.}
\begin{ruledtabular}
\begin{tabular}{|c|ccc|}
probability amplitude & ${\rm Re} \, c_{mn}$ & ${\rm Im} \, c_{mn}$  & $|c_{mn}|^2$ \\ \hline
$c_{00}$ & $-0.046$ & $0.016$ & $0.002$ \\
$c_{01}$ & $-0.061$ & $0.992$ & $0.987$ \\
$c_{10}$ & $0.049$ & $-0.007$ & $0.002$ \\
$c_{11}$ & $0.045$ & $ -0.030$ & $0.003$
\end{tabular}
\end{ruledtabular}
\vskip 0.2in
\end{table}

It is interesting to examine the extent to which higher energy states of the junction and resonator become excited during the storage operation. In Fig.~\ref{fig.storage-m} we plot the occupation probabilities of the states $|20\rangle$ and $|21\rangle$, both of which involve the higher lying $m \! = \! 2$ junction state. Similarly, in Fig.~\ref{fig.storage-n} we plot the occupations of $|02\rangle$ and $|12\rangle$, which involve the $n \! = \! 2$ phonon state. In all cases the excitation of higher lying states is negligible.

%\begin{figure}
%\includegraphics[width=.45\textwidth]{fig7.jpg}
%\caption[]{\label{fig.storage-m} Occupation of higher lying $m \! = \! 2$ junction state during qubit storage. The upper plot is $|c_{20}|^2 \! ,$ and the lower plot is $|c_{21}|^2 \! .$ Both quantities would vanish in the RWA.  All junction and resonator parameters are the same as in Fig.~\ref{fig.storage}.}
%\end{figure}

%\begin{figure}
%\vskip 0.2in
%\includegraphics[width=.45\textwidth]{fig8.jpg}
%\caption[]{\label{fig.storage-n} Occupation of higher lying $n \! = \! 2$ resonator state during qubit storage. The upper plot is $|c_{02}|^2 \! ,$ and the lower plot is  $|c_{12}|^2 ;$ both vanish in the RWA.  Parameters are the same as in Fig.~\ref{fig.storage}.}
%\vskip 0.3in
%\end{figure}

A few comments about these results are in order: The observed sensitivity to the shape of $s(t)$ can be understood by recalling that in the absence of any dissipation or decoherence, the RWA requires the qubit to be {\it exactly} in resonance with the nanomechanical resonator. Therefore it is necessary to bring the two systems into resonance as quickly as possible without violating adiabaticity. The power-law tails associated with an arctangent function, for example, lead to considerable deviations from the desired RWA behavior, as we demonstrate in Fig.~\ref{fig.arctan}. We expect this sensitivity to be present in real systems as well. We also found that the validity of the RWA requires $g$ to be considerably smaller than $\hbar \omega_0$. The ratio $g/\hbar \omega_0$ for the system simulated in Fig.~\ref{fig.storage} is 1\%. When the resonator disk radius $R$ is increased to $2.3 \, \mu {\rm m}$, $g/\hbar \omega_0$ is then only 10\%, but the RWA already fails considerably. This strong-coupling breakdown is demonstrated in Fig.~\ref{fig.storage-strong}. The resonant Rabi period in this case is $11.4 \, {\rm ns}$. Of course, the value of $\Delta t$ used in Figs.~\ref{fig.arctan} and \ref{fig.storage-strong} are consequences of the RWA analysis, and better fidelity could be obtained by choosing $\Delta t$ differently.

%\begin{figure}
%\includegraphics[width=.45\textwidth]{fig9.jpg}
%\caption[]{\label{fig.arctan} Qubit storage with arctangent bias-current profile. All system parameters are the same as in Fig.~\ref{fig.storage}. The numerical result for $|c_{10}|^2 \! ,$ shown as a solid descending curve, is entirely different than that predicted by the RWA (dashed curve), even though the difference between the $s(t)$ profiles used here and in Fig.~\ref{fig.storage} is small. The qubit state is not correctly stored in the resonator.}
%\vskip 0.3in
%\end{figure}

%\begin{figure}
%\includegraphics[width=.45\textwidth]{fig10.jpg}
%\caption[]{\label{fig.storage-strong} Qubit storage in larger resonator. Here we simulate qubit storage in a $15 \, {\rm GHz}$ resonator with $R = 2.3 \, \mu {\rm m}$, so that $g/\hbar \omega_0 = 0.10$. All other resonator and junction parameters are the same as in Fig.~\ref{fig.storage}. The solid descending curve is $|c_{10}|^2 \! ,$ and the dashed curve shows the desired RWA behavior. The solid ascending curve is $|c_{01}|^2 \! .$ The RWA breaks down here because of the stronger interaction strength. The dotted curve is $s(t)$. Qubit storage fails again.}
%\vskip 0.2in
%\end{figure}

Up to this point we have only discussed storage of the simple qubit state $|1\rangle$. Storing general qubit states of the form $\alpha |0\rangle + \beta |1\rangle$ follows similarly, although achieving high fidelity requires more care. The reason is that the ramping of up of $s(t)$ introduces phase errors into Eq.~(\ref{storage initial condition}), the ``inital'' amplitudes that get swapped. This can be circumvented to a considerable extent by choosing an optimum value of the  {\it off-resonant} bias current. In Fig.~\ref{fig.equator} we show results of the successful storage of the qubit states $2^{-{1 \over 2}}(|0\rangle+|1\rangle)$ and $2^{-{1 \over 2}}(|0\rangle+i|1\rangle),$ which are on the equator of the Bloch sphere, using $s=0.180$ when detuned from the resonator.

%\begin{figure}
%\includegraphics[width=.50\textwidth]{fig11.jpg}
%\caption[]{\label{fig.equator} Storage of qubit states on the equator of the Bloch sphere. (a) Here the initial state is $2^{-{1\over 2}}(|0\rangle_{\rm J} \! + \! |1\rangle_{\rm J}) \otimes |0\rangle_{\rm res}.$ The solid descending curve is the squared overlap with the interaction-representation state $2^{-{1\over 2}}(|00\rangle+|10\rangle),$  and the ascending curve is the occupation of $2^{-{1\over 2}}(|00\rangle+|01\rangle).$ The dotted curve is $s(t)$. (b) The initial state is $2^{-{1\over 2}}(|0\rangle_{\rm J} \! + \! i |1\rangle_{\rm J}) \otimes |0\rangle_{\rm res}.$ The descending and ascending curves are the occupations of $2^{-{1\over 2}}(|00\rangle+i|10\rangle)$ and $2^{-{1\over 2}}(|00\rangle+i|01\rangle),$ respectively.}
%\vskip 0.3in
%\end{figure}

Finally, in Fig.~\ref{fig.transfer}, we present results of simulations of {\it two} junctions coupled to a resonator. The JJs are the same as in Fig.~\ref{fig.storage}, but the resonator in this case has radius $R= 0.459 \, \mu{\rm m}.$ Because the upper gate is now split, $g = 0.620 \, {\rm \mu eV}$ for each JJ. The instantaneous eigenstates of the uncoupled system can be written as $|m_1 m_2 n\rangle$, where $m_1$ and $m_2$ are the eigenstates of the junctions and $n$ is the phonon number of the resonator. The phase qubit is first stored in the resonator, as described above, and is then passed to the second identical junction. The result is a transfer of the qubit state $|1\rangle$ from one JJ to another. Only half a Rabi period of resonance with the second JJ is needed for this transferred state; in general, one and a half periods are required. The probability amplitudes after the transfer are given in Table \ref{transfer table}.

%\begin{figure}
%\includegraphics[width=.50\textwidth]{fig12.jpg}
%\caption[]{\label{fig.transfer} Qubit transfer between two identical Josephson junctions. The descending solid curve is $|c_{100}(t)|^2 \! ,$ the probability for the first junction to be in the $m=1$ excited state, and the rest of the system to be in the ground state. The state of the first junction is stored in the resonator as in Fig.~\ref{fig.storage}, the peaked curve giving $|c_{001}(t)|^2 \! .$  The ascending curve is $|c_{010}(t)|^2 \! ,$ the probability for the second JJ to be in the excited state. The solid and dotted trapezoidal curves show the bias currents $s_1(t)$ and $s_2(t)$ on the two junctions.}
%\vskip 0.2in
%\end{figure}

\begin{table}
\caption{\label{transfer table}
Final probability amplitudes $c_{m_1 m_2 n}$ after transfering qubit state from one junction to another through the nanomechanical resonator. Transfer succeeds with a fidelity squared of better than $97\%$.}
\begin{ruledtabular}
\begin{tabular}{|c|ccc|}
probability amplitude & ${\rm Re} \, c_{m_1 m_2 n}$ & ${\rm Im} \, c_{m_1 m_2 n}$  & $|c_{m_1 m_2 n}|^2$ \\ \hline
$c_{001}$ & $ -0.075$ & $0.003$ & $0.006$ \\ 
$c_{010}$ & $0.591$ & $0.790$ & $0.974$ \\
$c_{100}$ & $0.023$ & $0.038$ & $0.002$ 
\end{tabular}
\end{ruledtabular}
\vskip 0.2in
\end{table}

\section{TWO-JUNCTION ENTANGLEMENT\label{entanglement section}}

The nanomechanical resonator can also be used to produce states where the JJs are entangled, but the resonator remains in its ground state, unentangled with the junctions.  We assume that two identical JJs are attached to the same split-gate resonator. The instantaneous eigenstates of the uncoupled system are written as $|m_1 m_2 n\rangle$, where $m_1$ and $m_2$ are the eigenstates of the junctions and $n$ is the phonon number of the resonator.

The foundations for this operation have already been explained in Sec.~\ref{qubit storage and transfer section}: According to Table \ref{pulse table}, we can prepare an entangled state of two JJs by bringing the first junction, previously prepared in the state $|1\rangle_{\rm J1}$, into resonance with the resonator for one {\it quarter} of a vacuum Rabi period, or $\Omega \, \Delta t = \pi/2,$ which produces the interaction-representation state $2^{-{1\over 2}}(|001\rangle + |100\rangle)$. The first JJ is now maximally enangled with the resonator, while the second junction is in the ground state. After bringing the second junction into resonance for half of a Rabi period, the state of the resonator and second junction are swapped, thereby ``passing'' the resonator's component of the entangled state to the second junction. After detuning the second junction, the system is then left in the interaction-representation state
\begin{equation}
{|100\rangle - |010\rangle \over \sqrt{2}}  =  {|10\rangle_{\rm J}  - |01\rangle_{\rm J} \over \sqrt{2}} \! \otimes \! |0\rangle_{\rm res} .
\label{entangled state}
\end{equation}
The two Josephson junctions have been prepared in the maximally entangled Bell state $2^{-{1\over 2}}(|10\rangle_{\rm J}  - |01\rangle_{\rm J})$. To produce the state $2^{-{1\over 2}}(|10\rangle_{\rm J} + |01\rangle_{\rm J})$, the $\Omega \, \Delta t = \pi $ swap pulse should be replaced with a $\Omega \, \Delta t = 3\pi $ swap pulse.

In Fig.~\ref{fig.entangle} we present the results of a simulation of entangled state preparation. The JJs are the same as in Fig.~\ref{fig.storage}, and the resonator has radius $R= 0.459 \, \mu{\rm m},$ resulting in an interaction strength of $g = 0.620 \, {\rm \mu eV}$ for each JJ. The desired entangled state is prepared with a squared fidelity of about 92\%.

%\begin{figure}
%\includegraphics[width=.45\textwidth]{fig13.jpg}
%\caption[]{\label{fig.entangle}Preparation of entangled Josephson junctions. The thick solid curve is the probability for the system to be found in the interaction-representation state $2^{-{1\over 2}}(|100\rangle - |010\rangle)$. The thin solid and dashed lines are $s_1(t)$ and $s_2(t)$, respectively.}
%\vskip 0.2in
%\end{figure}

\section{LARGE-SCALE QUANTUM CIRCUIT\label{network section}}

A strength of our architecture is scalability: By introducing additional {\it bus} junctions coupled to a pair of resonators, each resonator with a slightly different dilatational mode frequency, the quantum states of the resonators can be swapped. This makes it possible to construct a large JJ array, with all phase qubits coupled. We call this layout a ``hub-and-spoke'' network, an example of which is shown in Fig.~\ref{fig.largescale}. Each bus qubit ``spoke'' couples each adjacent resonator  ``hub," allowing a completely scalable geometry without intrinsic size limits.

%\begin{figure}
%\includegraphics[width=9.5cm]{fig14.jpg}
%\caption{\label{fig.largescale} Architecture for a large-scale JJ quantum computer. In addition to the junctions coupled to a single resonator, as in Fig.~\ref{fig.fourqubits}, here there are additional bus junctions for transferring states between different resonators.}
%\end{figure}

The Hamiltonian for an arbitrary large-scale quantum information processing circuit consisting of ${\cal M}$ phase qubits and ${\cal N}$ nanomechanical resonators is constructed as follows. Let $I \! = \! 1,2,\dots,{\cal N}$ label the resonators, which for simplicity we assume to lie in a two-dimensional plane, and let $J \! = \! 1,2,\dots,{\cal M}$ label the junctions. Typically there will be many more JJs than resonators. Each junction can couple to one or two resonators, subject to the constraint that a resonator can support on the order of 10 junctions, and that, for convenience, bus qubits should connect adjacent resonators. The Hamiltonian for such a quantum computer, ignoring state preparation, manipulation, and readout circuitry, as well as all environmental coupling, energy relaxation, and decoherence, is
\begin{eqnarray}
H_{\rm qc} &\equiv&  \sum_{\scriptscriptstyle I} \hbar \omega_{\scriptscriptstyle I} a_{\scriptscriptstyle I}^\dagger a_{\scriptscriptstyle I}
+ \sum_{{\scriptscriptstyle J}m} \epsilon_{{\scriptscriptstyle J}m} c_{{\scriptscriptstyle J}m}^\dagger c_{{\scriptscriptstyle J}m} \nonumber \\
&-& i \sum_{\scriptscriptstyle IJ}  \sum_{mm'} g_{\scriptscriptstyle IJ} \, (a_{\scriptscriptstyle I} - a_{\scriptscriptstyle I}^\dagger) \, x_{{\scriptscriptstyle J}mm'} \,c_{{\scriptscriptstyle J}m}^\dagger c_{{\scriptscriptstyle J}m'} .
\label{qc hamiltonian}
\end{eqnarray}
Here $\omega_{\scriptscriptstyle I}$ is the dilatational mode frequency of resonator $I$, $a_{\scriptscriptstyle I}^\dagger$ and $a_{\scriptscriptstyle I}$ are dilatational-mode phonon creation and annihilation operators satisfying $[a_{\scriptscriptstyle I} , a_{\scriptscriptstyle I'}^\dagger ] = \delta_{\scriptscriptstyle I I'},$ $\epsilon_{{\scriptscriptstyle J}m}$ is the spectrum of phase qubit $J$, and $ c_{{\scriptscriptstyle J}m}^\dagger$ and $c_{{\scriptscriptstyle J}m}$ are creation and annihilation operators (either bosonic or fermionic) for states $m$ in junction $J$. The matrix $g_{\scriptscriptstyle IJ}$ gives the interaction strength between resonator $I$ and junction $J$; bus junctions have nonzero $g_{\scriptscriptstyle IJ}$ for two values of $I$, computational  junctions will have only one nonzero element. In Eq.~(\ref{qc hamiltonian}) we have also neglected a small capacitive interaction between phase qubits connected to the same resonator.

As we will demonstrate in future work, the resonator can be used to mediate two-qubit quantum logic between phase qubits connected to that resonator. The quantum circuit of Fig.~\ref{fig.largescale} then allows quantum logic to be performed between any pair of computational qubits $J_1$ and $J_2$. This is accomplished by swapping the state stored in $J_2$ with a phase qubit $J_1^\prime$ attached to the same resonator as $J_1$, performing the logical operation on $J_1$ and $J_1^\prime$, and then reswapping $J_1^\prime$ and $J_2$. Any pair of computational qubits in Fig.~\ref{fig.largescale} can also be controllably entangled.

\section{DISCUSSION\label{discussion section}}

We have introduced a design for a scalable, solid-state quantum computing architecture based on the integration of nanoelectromechanical resonators with Josephson junction phase qubits. Quantum states prepared in a Josephson junction can be passed to the nanomechanical resonator and stored there, and then can be passed back to the original junction or transferred to another with high fidelity. The resonator can also be used to produce entangled states between a pair of Josephson junctions. Universal two-qubit quantum logic will be addressed in future work. The architecture is analogous to one or more few-level atoms in an electromagnetic cavity, and the junction-resonator complexes can assembled in a hub-and-spoke layout, resulting in a large-scale quantum circuit.

The calculations presented here have ignored all effects of dissipation and decoherence, with the assumption that the associated lifetimes are longer than a few hundred ns. This is not unreasonable given the current experimental situation. Nor have we attempted to perform the operations as fast as possible, and we expect there to be considerable room for improvement in both speed and fidelity.

Finally, we emphasize that many of our results will apply to other resonator- or oscillator-based qubit coupling methods.\cite{ShnirmanPRL97,MakhlinNat99,MooijSci99,MakhlinJLTP00,YouPRL02,Yukon02,SmirnovPre02,BlaisPRL03,PlastinaPRB03,ZhouPRA04,Buisson01,BlaisPRA04,WallraffNat04,GirvinPre03,MarquardtPRB01,HekkingPre02,ZhuPRA03} In particular, the promising design being developed at Yale,\cite{BlaisPRA04,WallraffNat04,GirvinPre03} using charge qubits coupled to superconducting transmission line resonators, is very similar to the architecture discussed here.

\section{ACKNOWLEDGMENTS}

It is a pleasure to thank Steve Lewis, Kelly Patton, Emily Pritchett, and Andrew Sornborger for useful discussions. MRG was supported by the National Science Foundation under CAREER Grant No.~DMR-0093217. ANC was supported by the DARPA/DMEA Center for Nanoscience Innovation for Defence.

\appendix

\section{QUANTUM MECHANICS OF THE PIEZOELECTRIC RESONATOR\label{resonator appendix}}

Here we quantize the vibrational dynamics of the piezoelectric resonator. In the quantum limit, the first term in Eq.~(\ref{general solution form}), which describes the background strain generated by the charge $\sigma(t)$, becomes trivially quantized: It gets multiplied by the identity operator. 

The quantization of the fluctuation term $\delta u(z,t)$ proceeds similarly to that of ordinary phonons, although we have to treat the zero-frequency ($n=0$) mode separately. First we construct a complete set of orthonormal eigenfunctions from Eq.~(\ref {fluctuation definition}), namely
\begin{equation}
f_n(z) \equiv \sqrt{2 - \delta_{n0} \over b} \, \cos(n \pi z/b), \ \ \ n=0,1,2, \cdots .
\end{equation}
These eigenfunctions can be shown to satisfy orthonormality
\begin{equation}
\int_0^b dz \, f^*_m(z) f_n(z) = \delta_{mn} 
\label{orthonormality identity}
\end{equation}
and completeness
\begin{equation}
\sum_{n=0}^\infty f_n^*(x) f_n(x') = \delta(x-x'),
\label{completeness identity}
\end{equation}
although in our case the $f_n(z)$ are purely real. 

\begin{widetext}

The quantized displacement-fluctuation field is given by
\begin{equation}
\delta u(z) =  f_0(z) \, z_0 +  \sum_{n=1}^\infty \sqrt{\hbar \over 2 \rho_{\rm lin} v k_n} \bigg( f_n(z) \, a_n + f^*_n(z) \, a_n^\dagger \bigg), 
\label{field expansion}
\end{equation}
and its associated momentum density $\Pi \equiv \rho_{\rm lin} \partial_t u$ is
\begin{equation}
\Pi(z) = f_0(z) \,  p_0 -i  \sum_{n=1}^\infty \sqrt{\hbar \rho_{\rm lin} v k_n \over 2} \bigg( f_n(z) \, a_n - f^*_n(z) \, a_n^\dagger \bigg),
\label{momentum expansion}
\end{equation}
where $k_n$ is defined in Eq.~(\ref{k definition}). Here $z_0$ is the $z$ component of the resonator center-of-mass coordinate operator, $p_0$ is the $z$ component of the center-of-mass momentum operator, and $[z_0,p_0] = i \hbar$. The $n=0$ term is excluded in the summations of Eqs.~(\ref{field expansion}) and (\ref{momentum expansion}) because the corresponding frequency $v k_n$ vanishes; its separate inclusion in the form given above will enable the use of the completeness relation (\ref{completeness identity}) in the analysis below. The $a_n$ and $a_n^\dagger$ are bosonic phonon annihilation and creation operators satisfying $[a_n, a_{n'}^\dagger] = \delta_{nn'}$. $\rho_{\rm lin} \equiv M_{\rm res} /b$ is the {\it linear} mass density of the cylindrical resonator, with $M_{\rm res}$ the resonator's mass. Using Eq.~(\ref{completeness identity}) it can be shown that
\begin{equation}
[u(z) , \Pi(z') ]=[\delta u(z) , \Pi(z') ] = i \hbar \delta(z-z'),
\end{equation}
as required.

The final expression for the quantized displacement field is therefore
\begin{equation}
u(z,t) = - {h_{33} \sigma(t) \over {\tilde c}_{33} } \, z +  f_0(z) \, z_0(t)
+  \sum_{n=1}^\infty \sqrt{\hbar \over 2 \rho_{\rm lin} v k_n} \bigg( f_n(z) \, a_n \, e^{- i v k_n t} + f^*_n(z) \, a_n^\dagger \, e^{i v k_n t}  \bigg), 
\label{total quantized field}
\end{equation}
where $z_0(t)$ is in the Heisenberg representation. If the Hamiltonian for the center-of-mass dynamics is $p_0^2/2M_{\rm res}$, then $z_0(t) = z_0 + (p_0/M_{\rm res})t.$ Note that the center-of-mass mode does not produce any strain and does not enter into our final results.

\end{widetext}

Using Eq.~(\ref{total quantized field}) leads to 
\begin{equation}
\delta U = - {2 \over b} \sum_{n \, {\rm odd}} \sqrt{\hbar \over M_{\rm res} v k_n} \bigg( a_n + a_n^\dagger \bigg).
\label{quantized delta U}
\end{equation}
If we include only the fundamental dilatational $(n=1)$ mode in Eq.~(\ref{quantized delta U}), we obtain (suppressing the subscript on the dilatational phonon operators)
\begin{equation}
\delta U \approx - {2 \ell_{\rm res} \over b} \, \big( a + a^\dagger \big),
\end{equation}
where $\ell_{\rm res} \equiv \sqrt{\hbar / M_{\rm res} \, \omega_0}$ is the characteristic size of quantum fluctuations in this mode, and where $\omega_0$ is the dilatational frequency defined in Eq.~(\ref{omega0 definition}).

Assuming a harmonic vibrational dynamics for the resonator, and ignoring the center-of-mass motion, the resonator Hamiltonian is 
\begin{equation}
H_{\rm res} = \sum_{n=1}^\infty \hbar v k_n (a_n^\dagger a_n + {\textstyle{1 \over 2}}).
\label{full resonator hamiltonian}
\end{equation}
Keeping only the $n=1$ dilatational mode, and dropping the additive c-number constant, leads to Eq.~(\ref{resonator hamiltonian}). Using Eq.~(\ref{full resonator hamiltonian}), we then obtain
\begin{equation}
\delta {\dot U} = {i \over \hbar} [H_{\rm res}, \delta U] =  {2 i \over b} \sum_{n \, {\rm odd}} \sqrt{\hbar v k_n \over M_{\rm res}} \big( a_n - a_n^\dagger \big).
\label{U dot}
\end{equation}
The $n=1$ term on the right-hand-side of Eq.~(\ref{U dot}), when inserted into Eq.~(\ref{classical interaction}), yields the interaction Hamiltonian of Eq.~(\ref{interaction hamiltonian}) with the coupling constant given in Eq.~(\ref{coupling strength formula}).

\bibliography{/Users/mgeller/Papers/bibliographies/MRGqc,/Users/mgeller/Papers/bibliographies/MRGpre,/Users/mgeller/Papers/bibliographies/MRGbooks,/Users/mgeller/Papers/bibliographies/MRGgroup,/Users/mgeller/Papers/bibliographies/MRGnano,qcnotes}

\begin{thebibliography}{76}
\bibitem[{ope()}]{operationsnote}
\bibinfo{note}{We note, however, that the number of operations $\tau_{\rm
  \varphi} \Omega_{\rm R} / 2 \pi$ that could be carried out at the Rabi
  frequency $\Omega_{\rm R}$ used in Ref.~[\onlinecite{YuSci02}] was at best of
  order 10.}
\end{thebibliography}

\newpage

\begin{widetext}

\begin{figure}
\caption[]{\label{fig.washboard}(main panel) Effective potential $U(\delta)$ for dimensionless bias current $s \equiv I_b/I_0$ equal to $0.1,$ plotted in units of $E_{\rm J}$. (inset) Equivalent-circuit model for a current-biased Josephson junction. A capacitance $C$ and resistance $R$ are in parallel with an ``ideal" Josephson element, represented by a cross and having critical current $I_0$. A bias current $I_b$ is driven through the circuit. }
\end{figure}

\begin{figure}
\caption[]{\label{fig.plasmabarrier} Barrier height and plasma frequency as a function of the dimensionless bias current $s$. Here $\Delta U_0 \equiv 2 E_{\rm J}$ is the barrier height at zero bias, and $\omega_{{\rm p}0}$ is the zero-bias plasma frequency defined in Eq.~(\ref{zero-bias frequency}).}
\end{figure}

\begin{figure}
\caption[]{\label{fig.currentbiasjj} Metastable potential well in the cubic limit, showing the barrier of height $\Delta U$ that separates the metastable states $| 0 \rangle$, $| 1 \rangle$, and $| 2 \rangle$, from the continuum. This figure applies to the case of bias currents $s$ just below 1. The lowest two states are separated in energy by $\Delta E$.}
\end{figure}

\begin{figure}
\caption[]{\label{fig.twoqubit} Two-qubit circuit diagram. The computational qubits are the two JJs in the center, shown as crossed boxes, each coupled to one side of the piezoelectric disk resonator. Each crossed box represents a real JJ, modeled by an ideal Josephson element in parallel with a resistor and capacitor. The current bias and readout circuits for each qubit circuit are shown on the left and right sides of the figure. Note that there is no direct electrical connection between the two qubits.}
\end{figure}

\begin{figure}
\caption{\label{fig.fourqubits} Four current-biased JJs coupled to a nanoelectromechanical resonator. Each junction is connected to a metallic plate on the surface of the resonator that covers about one quarter of the surface. Because we make use of the fundamental dilatational mode, which is spatially uniform in the plane of the resonator, the qubits are all equally well coupled to that mode.}
\end{figure}

\begin{figure}
\caption[]{\label{fig.storage}Phase qubit storage. The solid descending curve is $|c_{10}(t)|^2$, the interaction-representation occupation probability of the $|10\rangle$ state, calculated numerically for the junction of Ref.~[\onlinecite{MartinisPRL02}] coupled to the $15 \, {\rm GHz}$ piezoelectric resonator described in Table \ref{resonator table}. The dashed curve is the same quantity calculated from the analytic RWA results of Sec.~\ref{RWA section}. The solid ascending curve is $|c_{01}(t)|^2$. The dotted curve shows the time dependence of the dimensionless bias current $s(t)$, which is varied to bring the phase qubit in resonance with the resonator after $5 \, {\rm ns}$.  The Rabi period on resonance, when $s=0.546$, is $113.69 \, {\rm ns}$. After the storage operation, $|c_{10}|^2 = 0.002$ and $|c_{01}|^2 = 0.987$. The inset shows an enlarged view of $|c_{10}(t)|^2$ during the ramping up of $s(t)$.}
\end{figure}

\begin{figure}
\caption[]{\label{fig.storage-m} Occupation of higher lying $m \! = \! 2$ junction state during qubit storage. The upper plot is $|c_{20}|^2 \! ,$ and the lower plot is $|c_{21}|^2 \! .$ Both quantities would vanish in the RWA.  All junction and resonator parameters are the same as in Fig.~\ref{fig.storage}.}
\end{figure}

\begin{figure}
\caption[]{\label{fig.storage-n} Occupation of higher lying $n \! = \! 2$ resonator state during qubit storage. The upper plot is $|c_{02}|^2 \! ,$ and the lower plot is  $|c_{12}|^2 ;$ both vanish in the RWA.  Parameters are the same as in Fig.~\ref{fig.storage}.}
\end{figure}

\begin{figure}
\caption[]{\label{fig.arctan} Qubit storage with arctangent bias-current profile. All system parameters are the same as in Fig.~\ref{fig.storage}. The numerical result for $|c_{10}|^2 \! ,$ shown as a solid descending curve, is entirely different than that predicted by the RWA (dashed curve), even though the difference between the $s(t)$ profiles used here and in Fig.~\ref{fig.storage} is small. The qubit state is not correctly stored in the resonator.}
\end{figure}

\begin{figure}
\caption[]{\label{fig.storage-strong} Qubit storage in larger resonator. Here we simulate qubit storage in a $15 \, {\rm GHz}$ resonator with $R = 2.3 \, \mu {\rm m}$, so that $g/\hbar \omega_0 = 0.10$. All other resonator and junction parameters are the same as in Fig.~\ref{fig.storage}. The solid descending curve is $|c_{10}|^2 \! ,$ and the dashed curve shows the desired RWA behavior. The solid ascending curve is $|c_{01}|^2 \! .$ The RWA breaks down here because of the stronger interaction strength. The dotted curve is $s(t)$. Qubit storage fails again.}
\end{figure}

\begin{figure}
\caption[]{\label{fig.equator} Storage of qubit states on the equator of the Bloch sphere. (a) Here the initial state is $2^{-{1\over 2}}(|0\rangle_{\rm J} \! + \! |1\rangle_{\rm J}) \otimes |0\rangle_{\rm res}.$ The solid descending curve is the squared overlap with the interaction-representation state $2^{-{1\over 2}}(|00\rangle+|10\rangle),$  and the ascending curve is the occupation of $2^{-{1\over 2}}(|00\rangle+|01\rangle).$ The dotted curve is $s(t)$. (b) The initial state is $2^{-{1\over 2}}(|0\rangle_{\rm J} \! + \! i |1\rangle_{\rm J}) \otimes |0\rangle_{\rm res}.$ The descending and ascending curves are the occupations of $2^{-{1\over 2}}(|00\rangle+i|10\rangle)$ and $2^{-{1\over 2}}(|00\rangle+i|01\rangle),$ respectively.}
\end{figure}

\begin{figure}
\caption[]{\label{fig.transfer} Qubit transfer between two identical Josephson junctions. The descending solid curve is $|c_{100}(t)|^2 \! ,$ the probability for the first junction to be in the $m=1$ excited state, and the rest of the system to be in the ground state. The state of the first junction is stored in the resonator as in Fig.~\ref{fig.storage}, the peaked curve giving $|c_{001}(t)|^2 \! .$  The ascending curve is $|c_{010}(t)|^2 \! ,$ the probability for the second JJ to be in the excited state. The solid and dotted trapezoidal curves show the bias currents $s_1(t)$ and $s_2(t)$ on the two junctions.}
\end{figure}

\begin{figure}
\caption[]{\label{fig.entangle}Preparation of entangled Josephson junctions. The thick solid curve is the probability for the system to be found in the interaction-representation state $2^{-{1\over 2}}(|100\rangle - |010\rangle)$. The thin solid and dashed lines are $s_1(t)$ and $s_2(t)$, respectively.}
\end{figure}

\begin{figure}
\caption{\label{fig.largescale} Architecture for a large-scale JJ quantum computer. In addition to the junctions coupled to a single resonator, as in Fig.~\ref{fig.fourqubits}, here there are additional bus junctions for transferring states between different resonators.}
\end{figure}

\end{widetext}

\end{document}